\documentclass[submission, Phys]{SciPost}
\usepackage{graphicx,amsmath,amssymb,epstopdf,wasysym}
\usepackage[space]{grffile}
\usepackage{hyperref}
\usepackage[normalem]{ulem}
\usepackage{esint}
\usepackage{placeins}

\newcommand{\be}{\begin{equation}}
\newcommand{\ee}{\end{equation}}
\newcommand{\bea}{\begin{eqnarray}}
\newcommand{\eea}{\end{eqnarray}}
\newcommand{\ba}{\begin{eqnarray*}}
\newcommand{\ea}{\end{eqnarray*}}
\newcommand{\dagga}{{\phantom{\dagger}}}
\newcommand{\bR}{\mathbf{R}}

\newcommand{\bq}{\mathbf{q}}

\newcommand{\bk}{\mathbf{k}}

\newcommand{\bkp}{\mathbf{k{'}}}

\newcommand{\bx}{\mathbf{x}}
\newcommand{\by}{\mathbf{y}}

\newcommand{\br}{\mathbf{r}}
\newcommand{\Ima}{\text{Im}}

\newcommand{\dis}{\displaystyle}

\newcommand{\up}{\uparrow}
\newcommand{\tdot}{\mathbin{\,\ddot{\cdot}}\,}
\newcommand{\down}{\downarrow}
\newcommand{\fract}[2]{\frac{\dis #1}{\dis #2}}
\newcommand{\Tr}{\mathrm{Tr}}
\newcommand{\eqn}[1]{(\ref{#1})}

\newcommand{\bra}[1]{\langle #1 \mid}
\newcommand{\ket}[1]{\mid #1 \rangle}
\newenvironment{eqs}%
{\begin{equation} \begin{aligned}}%
{\end{aligned} \end{equation} }
\newcommand{\beal}{\begin{eqs}}
\newcommand{\eal}{\end{eqs}}
\newcommand{\bealn}{\beal\nonumber}
\newcommand{\bw}{\begin{widetext}}
\newcommand{\ew}{\end{widetext}}
\newcommand{\esp}[1]{\text{e}^{#1}}

\newcommand{\ep}{\epsilon}
\newcommand{\bd}[1]{{\boldsymbol{#1}}}

\newcommand{\coloneq}{\mathrel{\mathop:}\mathrel{\mkern-1.2mu}=}

\definecolor{NAcol}{RGB}{200,0,140}

\usepackage{lineno}
\begin{document}

\begin{center}{\Large \textbf{Non-Perturbative Resolution of Luttinger's Theorem Violations in Strongly Correlated Systems}}\end{center}

\begin{center}
Nazzareno Africani\textsuperscript{1$\star$}, Herbert E{\ss}l\textsuperscript{2$\dagger$}, 
Alessandro Toschi\textsuperscript{2} and Michele Fabrizio\textsuperscript{1}
\end{center}

\begin{center}
{\bf 1} International School for Advanced Studies (SISSA), Via Bonomea 265, 34136 Trieste, Italy
\\
{\bf 2} Institute of Solid State Physics, TU Wien, 1040 Vienna, Austria
\\[\baselineskip]
$\star$ \href{mailto:nafrican@sissa.it}{\small nafrican@sissa.it}\,,\quad
$\dagger$ \href{mailto:herbert.essl@tuwien.ac.at}{\small herbert.essl@tuwien.ac.at}
\end{center}

\begin{center}
\today
\end{center}


\section*{Abstract}
{\bf Many-body physics faces an apparent paradox concerning violations of Luttinger's theorem in strongly correlated systems. The theorem relates the electron number to the Luttinger count---the monodromy of the logarithm of the imaginary-frequency Green's function 
$\bd{G(i\epsilon)}$---and, at finite magnetic field, connects the Hall conductivity to the magnetic-field derivative of the Luttinger count via the Streda formula. Luttinger's theorem seemingly follows from particle-number conservation alone, without resorting to perturbation theory, and yet it is found to be violated. Here, we provide a non-perturbative solution to this paradox. We show that the Luttinger integral, which accounts for the mismatch between the particle number and the Luttinger count, is a pure zero-frequency boundary term, which the conventional proof implicitly discards. Hence, violations of Luttinger's theorem do not require a breakdown of the skeleton-diagram formalism, whose validity is independent of the convergence of the skeleton-diagram expansion to the exact solution. This advancement in understanding resolves a longstanding point of contention in many-body theory. We also find that the Luttinger integral is a genuine topological invariant, tied to the Luttinger-Ward functional rather than to the Green's function alone; like the monodromy of $\bd{\ln G(i\epsilon)}$, it stems exclusively from the zero-frequency point where the Matsubara trajectory crosses the branch cut of the logarithm. 
Correspondingly, the magnetic-field derivative of the Luttinger integral makes the Hall number in topological insulators differ from its Green's-function expression by a vertex correction, itself invariant, which can take fractional values.
}

\vspace{10pt}
\noindent\rule{\textwidth}{1pt}
\tableofcontents\thispagestyle{fancy}
\noindent\rule{\textwidth}{1pt}
\vspace{10pt}

\section{Introduction}
\label{Introduction}
Luttinger's theorem \cite{Luttinger} is a fundamental concept in modern theories of interacting electrons. Initially derived as a straightforward consequence of self-consistent many-body perturbation theory \cite{Luttinger}, its significance has evolved over time, with a growing emphasis on its topological nature \cite{Yunoki-PRB2017,Oshikawa-PRL2000,P2-Luttinger}, thereby transcending its perturbative origins. Indeed, it is related to one of the simplest topological quantities that can be constructed from the fully interacting single-particle Green's function. It also has a profound connection to another celebrated topological response, the quantum Hall conductivity, through the Streda formula, as recently pointed out by Gavensky, Sachdev and Goldman \cite{PhysRevLett.131.236601}.  \\

\noindent
At zero temperature, $T=0$, where the discrete Matsubara frequency $\ep_n=(2n+1)\pi T$ becomes a continuous variable $\ep \in[-\infty,\infty]$, the Luttinger count $N_L$ is defined by
\beal
N_L &\coloneq \sum_\bk\,\fint_{-\infty}^\infty \fract{d\ep}{2\pi}\;\esp{i\ep 0^+}\;
\fract{\partial}{\partial i\ep}\, \Tr\Big(\ln\, G(i\ep,\bk)^{-1}\Big)\,,
\label{Luttinger's theorem}
\eal  
where $G(i\ep,\bk)$ is 
the Green's function matrix represented in a generic basis of $M$ single-particle Bloch wavefunctions, spin included, with $\bk$ the lattice momentum, and $\fint$ the principal value at $\ep=0$, see Sec.~\ref{Sec2}. Luttinger's theorem asserts that the electron number $N$ at $T=0$ satisfies $N= N_L$.\\

\noindent
According to the Streda formula \cite{Streda_1982,Streda_1983}, 
the Hall conductivity at $T=0$ in units where $\hbar=1$ and, for instance, in two-dimensional insulators, can be obtained through  
\beal
\sigma_\text{H} &= -\fract{ec}{V}\;\fract{\partial N}{\partial B}\coloneq \fract{e^2}{2\pi}\, \nu_\text{H} \;,
\label{Streda}
\eal
where $V$ is the surface of the sample, $B$ a uniform magnetic field perpendicular to it, $-e<0$ the electron charge, and 
$\nu_\text{H}$ is the dimensionless Hall number. It follows that, if we define the finite-field counterpart of the Luttinger count as
\be
\sigma_\text{H\,L} \coloneq -\fract{ec}{V}\;\fract{\partial N_L}{\partial B}\coloneq \fract{e^2}{2\pi}\, \nu_\text{H\,L}\;,
\label{Streda 1}
\ee
then $\nu_\text{H}= \nu_\text{H\,L}$ if Luttinger's theorem \eqn{Luttinger's theorem} holds. 
It has been demonstrated \cite{PhysRevLett.131.236601,Andrea-PRB2023} that 
$\nu_\text{H\,L}$ coincides with 
the Ishikawa-Matsuyama-Volovik \cite{Ishikawa-1,Ishikawa-2,Volovik-2} expression for the Hall number
\be
\nu_\text{H\,L} =  \fract{i \pi}{V}\,\sum_\bk\,\int\fract{d\ep}{2\pi}\,\ep_{\mu\nu}\,
\Tr\Big(G(i\ep,\bk)\cdot\partial_\mu G(i\ep,\bk)^{-1}\cdot
\partial_{i\ep} G(i\ep,\bk)\cdot\partial_\nu G(i\ep,\bk)^{-1}\Big)\,,
\label{I and M 1}
\ee
where $\ep_{\mu\nu}$ is the antisymmetric tensor and $\partial_\mu\coloneq \partial_{k_\mu}$. The dimensionless quantity $\nu_\text{H\,L}$ is another Green's function
topological invariant, specifically, 
the winding number of the map $(\ep,\bk)\to G(i\ep,\bk)\in \text{GL}(M,\mathbb{C})$. 
It follows that $\nu_\text{H}= \nu_\text{H\,L}$ serves as yet another assertion of Luttinger's theorem. \\

\noindent
Although $N_L$ \cite{note-P2} and, consequently, $\nu_\text{H\,L}$ are legitimate  topological invariants, there is compelling evidence \cite{Altshuler-EPL1998,Rosh-2007,Phillips-PRL2013,Hewson-2018,PhysRevLett.90.216403,PhysRevB.70.245118,Slagle-PRB2015,Heath_2020,mio-3,Chubukov-LT} suggesting that in correlated metals and insulators Luttinger's theorem can be violated, thereby implying that $N \neq N_L$ and $\nu_\text{H} \neq \nu_\text{H\,L}$. 
A particularly transparent example is provided by fractional Chern insulators, where the physical $\nu_\text{H}$ is fractional, whereas $\nu_\text{H\,L}$ remains integer-valued \cite{Anton-Lucila-Nathan}.\\
As mentioned at the beginning, the Luttinger theorem was originally proved \cite{Luttinger}  exploiting the skeleton-diagram \emph{expansion} \cite{AG&D}, i.e.~assuming the validity of the self-consistent (or \textit{bold}) many-body perturbation theory.
Hence, the most natural explanation for possible violations of the Luttinger theorem is that its skeleton-diagram derivation \cite{Luttinger} ceases to hold when the skeleton-diagram expansion breaks down \cite{Altshuler-EPL1998,Schaefer2013,Kozik2015,Gunnarsson2017}. \\
The breakdown of perturbation theory, which underlies relevant physical features (such as the electronic localization preceding Mott-Hubbard metal-insulator transitions \cite{Chalupa2021,Mazitov2022,Pelz2023,Adler2024}, adjacent phase-separation instabilities \cite{Reitner2020,Reitner2024} or poles in the self-energy \cite{Phillips-PRL2013,Heath_2020,mio-2,Giorgio-NatComm2023,mio-3,mio-Mott,Chubukov-LT}), is directly linked to important formal consequences in the many-body theory (such as divergences of the two-particle irreducible vertices \cite{Schaefer2013,Janis2014,Toschi-2018,Vucicevic2018,Springer2020}, crossings of solutions of the Luttinger-Ward functional \cite{Kozik2015,Rossi2015,Stan2015,Gunnarsson2017,Kim2022}, and misleading convergences of the iterative diagrammatic algorithms \cite{Essl2026,Esslpreprint26,Gieverspreprint26}).\\ 

\noindent
At the same time, we must emphasize that the Luttinger theorem can also be derived fully \emph{non-perturbatively} \cite{Potthoff-2006}. Therefore, a breakdown of the skeleton-diagram expansion is \emph{not} sufficient, per se, to account for the violation of the theorem (for the sake of clarity, throughout this paper we make a distinction between the skeleton-diagram expansion and the skeleton-diagram formalism~\cite{footnote}). \\
In this respect, we note that, if one excludes the spurious/unphysical solutions \cite{Kozik2015,Rossi2015,Gunnarsson2017} of the (multivalued) LW functional, all dynamical properties of the interacting systems, which are dependent on one frequency (such as, e.g., the one-particle Green's function or the Kubo response functions) are well behaved for any nonzero imaginary frequency \cite{Igor-PRB2003}, since the only physically meaningful discontinuities in these quantities can occur at \emph{zero} Matsubara frequency. There, the imaginary frequency axis crosses the real axis, where the branch cut of the physical Green's (or Kubo response) function  lies.
For instance, $\Tr\big(G(i0^-,\bk)\big)-\Tr\big(G(i0^+,\bk)\big)$ coincides with $2\pi i$ times the momentum-dependent single-particle density of states at the chemical potential, which is finite in a metal. \\

\noindent
Building upon this insight, we demonstrate that the occurrence of Luttinger's theorem violations is exclusively associated with the above mentioned discontinuities at $\ep=0$. Consequently, \emph{these} discontinuities must be handled with utmost care under a fixed branch prescription. Indeed, as we show below, they generate a nontrivial global contribution in the Green's function space associated with the Luttinger-Ward functional.\\

\noindent
Specifically, we demonstrate in Secs.~\ref{A more in-depth examination} and 
\ref{The Luttinger integral is a pure boundary term} that the so-called Luttinger integral $I_L\coloneq N-N_L$, which takes integer values at zero temperature, is a pure $\ep=0$ boundary term. This implies that the conventional proof of Luttinger's theorem, as presented in Sec.~\ref{Sec2}, has a flaw because it fails to account for such a boundary term that cannot be assumed to vanish a priori, 
as we discuss thoroughly in Secs.~\ref{Paradox of the functional chain rule and flaw in theorem's proof} and \ref{The paradox in Matsubara formalism}.
In other words, the conventional proof, both in its original perturbative form~\cite{Luttinger} and in its non-perturbative formulation~\cite{Potthoff-2006}, implicitly discards this boundary term. Since $I_L$ is itself that boundary term, this amounts to assuming $I_L=0$, i.e. the very statement of the theorem. Therefore, also the arguments used to justify that $\nu_\text{H}=\nu_\text{H\,L}$ do not hold in general, and the mismatch $\nu^I_\text{H\,L}=\nu_\text{H}-\nu_\text{H\,L}$, derived explicitly in Sec.~\ref{Sec3}, can be finite when Luttinger's theorem is violated. \\
The above results raise a relevant question: Which conditions do then guarantee the validity of the theorem in situations where it is found to hold? 
We demonstrate that the Luttinger integral $I_L$ is a genuine topological invariant, see Sec.~\ref{The Luttinger Integral as a novel topological invariant}. Similarly, the magnetic field analogue $\nu^I_\text{H\,L}$ is also a topological invariant, as shown in Appendix \ref{Invariance property}. Consequently, if $I_L=0$ for a specific point in the Hamiltonian parameter space, it must maintain that value within a finite region surrounding that point. In particular, since $I_L$ trivially vanishes in the absence of interaction, Luttinger's theorem must hold also for finite but weak enough interaction. \\ 
In turn, the topological nature of the Luttinger integral raises another pertinent question: How can the value of $I_L$ change? We find that this occurs through a singularity of the self-energy at $\ep=0$, see Sec.~\ref{The Luttinger Integral as a novel topological invariant}. In a periodic model, this corresponds to the emergence of a so-called Luttinger surface \cite{Igor-PRB2003}, i.e. the location in the Brillouin zone of the zeros of 
$\text{det}\big(G(0,\bk)\big)$. As long as the Luttinger surface exists, $I_L$ is not topologically protected and may vary. However, when the Luttinger surface disappears, $I_L$ has changed into a new quantized value and the self-energy becomes again analytic also at $\ep=0$. In finite systems, the self-energy becomes singular at isolated 
points in the Hamiltonian parameter space, after which $I_L$ changes value, and the self-energy regains full analytic behavior, as we explicitly show in a few simple examples in Sec.~\ref{A few relevant examples}.

\section{The Luttinger theorem}
\label{Sec2}
Since Luttinger's theorem does not rely on any symmetry but particle conservation, we find it convenient in this section not to specify the basis of the single-particle wavefunctions $\phi_\alpha(\br)$, 
$\alpha=1,\dots,L$ that includes the spin ($L=MV$ for $V$ unit cells). The thermal Green's function is therefore an $L\times L$ 
complex matrix satisfying $G(i\ep_n)=G(-i\ep_n)^\dagger$, with matrix elements $G_{\alpha\beta}(i\ep_n)$ and $\ep_n=(2n+1)\pi T$ the Matsubara frequency. We can rewrite $G(i\ep_n)$ as 
\be
G(i\ep_n) = G_1(\ep_n)+ i\,G_2(\ep_n)\,,
\label{LT: G1 and G2}
\ee
where $G_1(\ep_n)=G_1(\ep_n)^\dagger=G_1(-\ep_n)$, while 
$G_2(\ep_n)=G_2(\ep_n)^\dagger=-G_2(-\ep_n)$, which is negative definite for $\ep_n>0$ and positive definite otherwise. \\
If $\Phi[G]$ is the Luttinger-Ward (LW) functional, then the self-energy 
$\Sigma(i\ep_n) = \Sigma(-i\ep_n)^\dagger$ has matrix elements defined through 
\beal
\Sigma_{\alpha\beta}(i\ep_n) = 
\fract{\delta\Phi[G]}
{\delta G_{\beta\alpha}(i\ep_n)} \;,
\label{self energy}
\eal
and can be also written as  
\be
\Sigma(i\ep_n) = \Sigma_1(\ep_n) + i\,\Sigma_2(\ep_n)\,,
\label{LT: Sigma1 and Sigma2}
\ee
where $\Sigma_1(\ep_n)$ and $\Sigma_2(\ep_n)$ have the same properties as $G_1(\ep_n)$ and  
$G_2(\ep_n)$ in \eqn{LT: G1 and G2}. Self-energy and Green's function 
are related to each other by the Dyson equation
\beal
G(i\ep_n)^{-1} &= i\ep_n -H_0 -\Sigma(i\ep_n)\,,
\label{Dyson}
\eal
where $H_0$ is the non-interacting Hamiltonian represented in the chosen basis and which includes the chemical potential term. \\
Since the number of electrons $N$ is a conserved quantity, the LW functional is invariant under a shift of the Matsubara frequencies of all internal Green's functions by the same bosonic frequency $\omega_m=2m\pi T$.   
This invariance property can be expressed mathematically as:  
\beal
\Phi\big[G(i\ep_n+i\omega_m)\big] -\Phi\big[G(i\ep_n-i\omega_m)\big]\equiv 0\,,& 
&\forall \,\omega_m=2m\pi T\,,
\label{LT invariance}
\eal
where we use a symmetrized difference for simplicity. 
For $T\to 0$ also $\omega_m\to 0$ at fixed $m$, which formally enables us to rewrite \eqn{LT invariance} using \eqn{self energy} as
\beal
0 &\equiv \lim_{T\to 0}\;\fract{\;\Phi\big[G(i\ep_n+i\omega_m)\big] -\Phi\big[G(i\ep_n-i\omega_m)\big]\;}{2i\omega_m}\\
& \overset{?}{\equiv} \lim_{T\to 0}\; T\sum_n\,
\esp{i\ep_n 0^+}\;
\Tr\bigg(\Sigma(i\ep_n)\;\fract{\;G(i\ep_n+i\omega_m)-G(i\ep_n-i\omega_m)\;}{2i\omega_m}\bigg)
\\
&\quad =\lim_{T\to 0}\; T\sum_n\,\Tr\bigg(\Sigma(i\ep_n)\;\fract{\;G(i\ep_n+i\omega_m)-G(i\ep_n-i\omega_m)\;}{2i\omega_m}\bigg)
\\
&\quad =-\lim_{T\to 0}\; T\sum_n\,\Tr\bigg(G(i\ep_n)\;\fract{\;\Sigma(i\ep_n+i\omega_m)-\Sigma(i\ep_n-i\omega_m)\;}{2i\omega_m}\bigg)\,.
\label{LT invariance T=0 0}
\eal
We have marked the second equality with a question mark because, as we show in Sec.~\ref{Paradox of the functional chain rule and flaw in theorem's proof}, this superficially innocuous application of the functional chain rule is precisely where the conventional proof of Luttinger's theorem fails. We adopt this convention throughout, to flag every subsequent occurrence of the same step.
Moreover, we disregard the regularizing factor $\esp{i\ep_n 0^+}$, which is permissible since the sums converge absolutely at infinity. \\
Let us analyze separately the $T\to 0$ limits of the two sums in \eqn{LT invariance T=0 0}. 
We first note that $G(i\ep_n+i\omega_m)-G(i\ep_n-i\omega_m)$ may not vanish for $T\to 0$ if $\ep_n+\omega_m > 0$ but $\ep_n-\omega_m < 0$ for $\omega_m>0$. Taking, for simplicity, $\omega_m=\omega_1=2\pi T$, 
this condition holds if $n=-1,0$, so that, using \eqn{LT: G1 and G2} and \eqn{LT: Sigma1 and Sigma2},   
\beal
&T\sum_n\,\Tr\Bigg(\Sigma(i\ep_n)\;\fract{\; G(i\ep_n+i\omega_1)-G(i\ep_n-i\omega_1)\;}{2i\omega_1}\Bigg)\\
 &\qquad = \fract{1}{4\pi i}\, \bigg\{
 \Tr\Big(\Sigma(i\pi T)\;\big(G(3i\pi T)-G(-i\pi T)\big)\Big)\\
 &\qquad\qquad \qquad
 +\Tr\Big(\Sigma(-i\pi T)\;\big(G(i\pi T)-G(-3i\pi T)\big)\Big)\bigg\}
 \\
 &\qquad \quad + T\sum_{n\not=-1,0}\,\Tr\Bigg(\Sigma(i\ep_n)\;\fract{\ G(i\ep_n+i\omega_1)-G(i\ep_n-i\omega_1)\;}{2i\omega_1}\Bigg)\\
&\qquad \xrightarrow[T\to 0]{} \, \fract{1}{\pi}\,\Tr\Big(\Sigma_1(0^+)\,G_2(0^+)\Big)
+ \fint_{-\infty}^\infty \fract{d\ep}{2\pi}\;
\Tr\Big(\Sigma(i\ep)\, \partial_{i\ep} G(i\ep)\Big)\,,
\label{Sigma partial G}
\eal
where $\fint$ is the principal value at $\ep=0$ and $\int$ the integral in the sense of distributions. At $T=0$ a Matsubara sum $T\sum_n f(i\ep_n)$ becomes a $\fint$, the fermionic frequencies being symmetric around $\ep=0$, whereas a sum of finite differences, e.g.\ $T\sum_n\big[F(i\ep_n+i\omega_m)-F(i\ep_n-i\omega_m)\big]/2i\omega_m$, becomes a $\int$: the principal value plus the jump at $\ep=0$, which in \eqn{Sigma partial G} is the boundary term. The latter vanishes for an insulator since $G_2(0^+)\equiv 0$, but is finite in a metal, $G_2(0^+)<0$.  Similarly, 
\beal
&T\sum_n\,\Tr\Bigg(G(i\ep_n)\;\fract{\; \Sigma(i\ep_n+i\omega_1)-\Sigma(i\ep_n-i\omega_1)\;}{2i\omega_1}\Bigg)\\
&\qquad \xrightarrow[T\to 0]{} \, \fract{1}{\pi}\,\Tr\Big(G_1(0^+)\,\Sigma_2(0^+)\Big)
+ \fint_{-\infty}^\infty \fract{d\ep}{2\pi}\;
\Tr\Big(G(i\ep)\, \partial_{i\ep} \Sigma(i\ep)\Big)\,.
\label{G partial Sigma}
\eal
Now, the boundary term vanishes whenever 
\be
\Tr\Big(G_1(0^+)\,\Sigma_2(0^+)\Big) = 0\,,
\label{main assumption}
\ee
which we henceforth assume and encompasses both metals and insulators with or without a Luttinger surface. Indeed, in the presence of a Luttinger surface, even though $\Sigma_2(0^+)$ is singular, \eqn{main assumption} still holds. Models where 
\eqn{main assumption} does not hold include, for instance, the SYK model \cite{Georges-PRB2001,Sachdev-RMP2022}. In conclusion, we can write \eqn{LT invariance T=0 0} as
\beal
0 &\equiv \lim_{T\to 0}\;\fract{\;\Phi\big[G(i\ep_n+i\omega_m)\big] -\Phi\big[G(i\ep_n-i\omega_m)\big]\;}{2i\omega_m}\\
& \overset{?}{\equiv} 
\fract{1}{\pi}\,\Tr\Big(\Sigma_1(0^+)\,G_2(0^+)\Big)
+ \fint_{-\infty}^\infty \fract{d\ep}{2\pi}\;
\Tr\Big(\Sigma(i\ep)\, \partial_{i\ep} G(i\ep)\Big)\\
& \qquad = -\fint_{-\infty}^\infty \fract{d\ep}{2\pi}\;
\Tr\Big(G(i\ep)\, \partial_{i\ep} \Sigma(i\ep)\Big)
\,.
\label{LT invariance T=0}
\eal
\noindent
The electron number $N$ can be obtained through 
\be
N = T\,\sum_n\,\esp{i\ep_n 0^+}\;\Tr\Big(G(i\ep_n)\Big)\,,
\label{LT def N}
\ee
where the regularizing exponential factor is required since $G(i\ep_n)\sim 1/i\ep_n$ 
for large $|\ep_n|$. For $T\to 0$, $N$ becomes an integer, the sum over the Matsubara frequency an integral, so that \eqn{LT def N} can be written through \eqn{Dyson} as \cite{Luttinger}
\beal
N &= \fint_{-\infty}^\infty \fract{d\ep}{2\pi}\;\esp{i\ep 0^+}\;\Tr\Big(G(i\ep)\Big)\\
&= \fint_{-\infty}^\infty \fract{d\ep}{2\pi}\;\esp{i\ep 0^+}\;\fract{\partial}{\partial i\ep}\,
\Tr\Big(\ln G(i\ep)^{-1}\Big)
+ \fint_{-\infty}^\infty \fract{d\ep}{2\pi}\;\esp{i\ep 0^+}\;
\Tr\Bigg(G(i\ep)\;\fract{\partial \Sigma(i\ep)}{\partial i\ep}\Bigg)\\
&\coloneq N_L + I_L\,,
\label{LT N T=0}
\eal
which serves as a definition of the Luttinger integral $I_L$, and $N_L$ is the Luttinger count defined in Eq.~\eqn{Luttinger's theorem}. As discussed above, we can safely ignore the regularizing exponential factor in $I_L$. Comparing \eqn{LT N T=0} with \eqn{LT invariance T=0} one should conclude that the Luttinger integral vanishes identically, thus that $N= N_L$ at $T=0$. This is precisely the statement of Luttinger's theorem \cite{Luttinger}.  \\
Except for the step marked with a question mark, the derivation is free of any apparent flaw and extends beyond the validity limits of perturbation theory, as discussed by Potthoff in \cite{Potthoff-2006}. Given the several known violations of Luttinger's theorem, it must be precisely that step which fails, as we show in the following.

\subsection{A more in-depth examination}
\label{A more in-depth examination}
Let us begin our analysis by exploring more in detail the properties of the Green's function $G(i\ep)$ and of the Luttinger count. We define, using Eq.~\eqn{LT: Sigma1 and Sigma2}, the \textit{quasiparticle} residue matrix \cite{Andrea-PRB2023} as
\beal
Z(\ep) &= Z(-\ep) = \Bigg(1 - \fract{\;\Sigma_2(\ep)\;}{\ep}\Bigg)^{-1}\,,
\label{Z(epsilon)}
\eal
which is  positive semi-definite with eigenvalues $\in[0,1]$. Through \eqn{Z(epsilon)} we can write 
\beal
G(i\ep) &= \sqrt{Z(\ep)\;}\;\fract{1}{\;i\ep - H_*(\ep)\;}\;\sqrt{Z(\ep)\;}
\coloneq \sqrt{Z(\ep)\;}\;G_*(i\ep)\;\sqrt{Z(\ep)\;}\;,
\label{G qp}
\eal
where 
\be
H_*(\ep) = H_*(\ep)^\dagger = H_*(-\ep) = 
\sqrt{Z(\ep)\;}\;\Big(H_0 + \Sigma_1(\ep)\Big)\;\sqrt{Z(\ep)\;}\;,
\label{H_*}
\ee
is the \textit{quasiparticle} Hamiltonian with eigenvalues $\ep_\ell(\ep)=\ep_\ell(-\ep)$ 
and eigenstates $\ket{\!\!u_\ell(\ep)}$, $\ell=1,\dots,L$, which depend parametrically on $\ep$. We denote 
$\ep_\ell(\ep=0)=\ep_\ell$ as the actual \textit{quasiparticle} energies. Since 
$Z(\ep)=Z(-\ep)\simeq 1-O(1/\ep^2)$ for large $|\ep|$, one readily finds that 
\beal
N_L &=  \fint_{-\infty}^\infty \fract{d\ep}{2\pi}\;\esp{i\ep 0^+}\;\fract{\partial}{\partial i\ep}\,
\Tr\Big(\ln G(i\ep)^{-1}\Big) = \fint_{-\infty}^\infty \fract{d\ep}{2\pi}\;\esp{i\ep 0^+}\;\fract{\partial}{\partial i\ep}\,
\Tr\Big(\ln G_*(i\ep)^{-1}\Big)\\
&= \sum_\ell\, \theta\big(-\ep_\ell\big)\coloneq \sum_\ell\, n_{L\ell}\,,
\label{Luttinger count 0}
\eal 
which also clarifies the term \textit{quasiparticles} that we adopted, and where 
the Luttinger occupation number $n_{L\ell}$ is either 1 or 0.\\
We represent the matrix $G(i\ep)$ in the basis of the 
eigenstates $\ket{\!u_\ell(0)}$ of $H_*(\ep=0)$. For each eigenstate, we associate the annihilation operator $c^\dagga_\ell$ and the creation operator $c^\dagger_\ell$. In this basis, the diagonal elements $G_{\ell\ell}(i\ep)$ of the Green's function $G(i\ep)$ have the following spectral decomposition at $T=0$:
\be
G_{\ell\ell}(i\ep) = 
\sum_{n>0}\,\Bigg\{
\fract{\;\big|\bra{n}c^\dagger_{\ell}\ket{0}\big|^2\;}
{i\ep-E_n+E_0}
+ \fract{\;\big|\bra{n}c^\dagga_{\ell}\ket{0}\big|^2\;}
{i\ep+E_n-E_0}\Bigg\}
\coloneq \sum_{n>0}\,\Bigg\{
\fract{\;R^+_{\ell n}\;}
{\;i\ep-E_{n0}\;}
+ \fract{\;R^-_{\ell n}\;}
{\;i\ep+E_{n0}\;}\Bigg\}\,,
\label{G diag}
\ee
where $\ket{\!n}$ are the many-body eigenstates with eigenvalues $E_n$, $n\geq 0$, 
with $\ket{\!0}$ the ground state, assumed non-degenerate, and $E_{n0}=E_n-E_0>0$ for $n>0$. 
By definition, 
\be
\sum_{n>0}\,R^+_{\ell n} = \bra{0}c^\dagga_{\ell}\,c^\dagger_{\ell}\ket{0}
\coloneq 1-n_\ell\in[0,1]\,,\quad 
\sum_{n>0}\,R^-_{\ell n} = \bra{0}c^\dagger_{\ell}\,c^\dagga_{\ell}\ket{0}
\coloneq n_\ell\in[0,1]\,.
\label{sum R}
\ee
Hereafter, besides \eqn{main assumption},
 we perform our analysis under the following conditions:
 (i) we assume, as mentioned above, that the ground state is \emph{non-degenerate}; and (ii) we discard sets of measure zero in the Hamiltonian parameter space where the two-particle irreducible vertex function is indeterminate \cite{Schaefer2013,Janis2014,Toschi-2018,Vucicevic2018,Springer2020}.\\
Assumption (i) rules out possible spurious effects on $I_L$ directly linked to the special situation of a degenerate many-body ground state. Assumption (ii) excludes pathological Hamiltonian parameter combinations where the skeleton diagrammatic formalism is locally ill-defined, despite being well-defined anywhere else away from such isolated points.\\
In the chosen basis of single-particle wavefunctions $\ket{\!u_\ell(0)}$, we can thus write the Luttinger integral \eqn{LT N T=0} simply as
\be
I_L \equiv N - N_L = \sum_\ell\,\Big(n_\ell - n_{L\ell}\Big)
= \sum_\ell\,\Big(n_\ell - \theta\big(-\ep_\ell\big)\Big)\,.
\label{true IL}
\ee
The equation above already explains how and why Luttinger's theorem can be violated. While $n_{L\ell}$ is always either 0 or 1, the actual $n_\ell$ can be any number in the range $[0,1]$. This deviation occurs when part of the spectral weight is transferred across the chemical potential \cite{Haldane-LT}. In a weakly correlated insulator, different deviations compensate exactly to yield $I_L=0$; thus, Luttinger's theorem is valid. However, when correlation is strong, single bands split into lower and upper Hubbard bands, with the lower band potentially accommodating any fractional number of electrons.
In this case, some of the \textit{quasiparticle} bands lose their connections with poles of the physical single-particle spectrum \cite{Andrea-PRB2023}, and the perfect compensation stops, leading to a finite $I_L$ and the violation of Luttinger's theorem.

\subsection{The Luttinger integral is a pure boundary term}
\label{The Luttinger integral is a pure boundary term}

However, the above observations explain why $I_L$ can become non-zero, but do not explain where the theorem's proof fails. To address this, under the assumption of a fixed branch cut of the logarithm, i.e. 
$\Ima \ln z = \Ima \ln (x+iy) = \text{Arg}(z) \in[-\pi,\pi]$, we rewrite \eqn{G diag} as 
\beal
G_{\ell\ell}(i\ep) &= \fract{\partial}{\partial i\ep}\,
\sum_{n>0}\,\Big\{ R^+_{\ell n}\,\ln\big(i\ep-E_{n0}\big)
+ R^-_{\ell n}\,\ln\big(i\ep+E_{n0}\big)\Big\}\\
&= \fract{\partial}{\partial i\ep}\, \ln\Bigg(\prod_{n>0}\,
\big(i\ep-E_{n0}\big)^{R^+_{\ell n}}\;\big(i\ep+E_{n0}\big)^{R^-_{\ell n}}\Bigg)
\coloneq \fract{\partial \ln S_\ell(i\ep)}{\partial i\ep}\,,
\label{G diag 1}
\eal
where we consider $S_\ell(i\ep)$ the elements of a diagonal matrix $S(i\ep)$. 
Writing the sum of the logarithms as the logarithm of the product is consistent with the 
fixed branch cut assumption. Indeed, since $0\leq R^\pm_{\ell n}\leq 1$ and, through 
\eqn{sum R}, $\sum_{n>0}\, (R^+_{\ell n} + R^-_{\ell n}) = 1$, then  
\beal
\Ima\, \ln S_\ell(i\ep)&= - \Ima\, \ln S_\ell(-i\ep)\\
&=\sum_{n>0}\,\Bigg\{ 
R^+_{\ell n}\;\text{Arg}(i\ep-E_{n0})
+ R^-_{\ell n}\;\text{Arg}(i\ep+E_{n0})\Bigg\}\in 
\begin{cases}
[0,\pi] & \ep > 0\,,\\
[-\pi,0] & \ep<0\,,
\end{cases}
\label{Arg S ell}
\eal 
which ensures that the sum of the imaginary part of the logarithms does not fall outside the interval $[-\pi,\pi]$. 
Specifically,
\bealn
\Ima\, \ln S_\ell(\pm i0)&=\pm\pi\,\big(1-n_\ell\big)\,,&
\Ima\, \ln S_\ell(\pm i\infty)&= \pm \fract{\pi}{2}\;.
\eal
By the definition \eqn{LT N T=0} of the Luttinger integral we can conclude that 
\beal
I_L &= N-N_L = \fint_{-\infty}^\infty 
\fract{d\ep}{2\pi}\;\esp{i\ep 0^+}\;\fract{\partial}{\partial i\ep}\,
\bigg\{
\Tr\Big(\ln \,S(i\ep)\Big) - \Tr\Big(\ln \,G(i\ep)^{-1}\Big)\bigg\}\\
&= \fint_{-\infty}^\infty
\fract{d\ep}{2\pi}\;\fract{\partial}{\partial i\ep}\,
\bigg\{
\Tr\Big(\ln\, S(i\ep)\Big) - \Tr\Big(\ln\, G(i\ep)^{-1}\Big)\bigg\}
\,,
\label{IL start}
\eal
where we have ignored the regularizing exponential term since the integrand vanishes fast
at $\ep=\pm\infty$. In essence, $I_L$ is a pure $\ep=0$ boundary term, and this is the primary reason why the conventional proof of Luttinger's theorem is flawed. \\
For $\ep>0$, the anti-Hermitian part of $G(i\ep)^{-1}$, i.e. $\ep-\Sigma_2(\ep)$, is positive definite. Therefore, all eigenvalues of $G(i\ep)^{-1}$ lie in the upper half-plane, and those of $G(-i\ep)^{-1}=G(i\ep)^{-1\,\dagger}$ in the lower one, so that $\Ima\,\Tr\big(\ln G(i\ep)^{-1}\big)=-\Ima\,\Tr\big(\ln G(-i\ep)^{-1}\big)\in(0,L\pi)$ for $\ep>0$ never crosses a branch cut. By \eqn{Arg S ell}, the same holds for $\Tr\big(\ln S(i\ep)\big)$.
It follows that, given the prescription of a fixed branch cut, we are allowed to define
\be
\Tr\Big(\ln X(i\ep)\Big) = \Tr\Big(\ln S(i\ep)\Big) - \Tr\Big(\ln G(i\ep)^{-1}\Big) \,.
\label{def X}
\ee
Through the definition of $S(i\ep)$ in Eq.~\eqn{G diag 1} and the Dyson equation \eqn{Dyson} 
we find that, at any $\ep\not=0$, 
\be
\fract{\partial}{\partial i\ep}\,\Tr\Big(\ln X(i\ep)\Big)
=\Tr\Big(G(i\ep)\,\partial_{i\ep}\Sigma(i\ep)\Big)\,,
\label{pointwise}
\ee
thus
\beal
I_L &= \fint_{-\infty}^\infty
\fract{d\ep}{2\pi}\;\fract{\partial}{\partial i\ep}\,
\Tr\Big(\ln  X(i\ep)\Big) = -\fract{1}{\pi}\,
\Ima\,\Tr\Big(\ln X(i0^+)\Big)\,.
\label{I_L vs X}
\eal
For $\ep\to 0^+$, 
\be\nonumber
\Ima\,\Tr\Big(\ln S(i0^+)\Big)=\pi\sum_\ell\,(1-n_\ell)=\pi\,(L-N)\,,\quad 
\Ima\,\Tr\Big(\ln G(i0^+)^{-1}\Big)=\pi\,(L-N_L)\,,
\ee
the latter since $G(0)^{-1}=-Z(0)^{-1/2}\,H_*(0)\,Z(0)^{-1/2}$ is Hermitian and, by Sylvester's law of inertia, has as many negative eigenvalues as $H_*(0)$ has positive ones, i.e. $L-N_L$. Therefore,
\be
I_L = -\fract{1}{\pi}\,
\Ima\,\Tr\Big(\ln X(i0^+)\Big) = N-N_L \coloneq Z \in \mathbb{Z}\,.
\label{true IL bis}
\ee
since both $N$ and $N_L$ are integers at $T=0$. 
   
\subsection{Connections with the Luttinger-Ward functional}
\label{Connections with the Luttinger-Ward functional}

We can delve deeper into the previous results to finally pinpoint where the proof of Luttinger's theorem encounters a flaw. We note that Eq.~\eqn{pointwise}, together with the definition \eqn{self energy} of the self-energy, implies that
\beal
I_L &= \fint_{-\infty}^\infty \fract{d\ep}{2\pi}\;
\Tr\Bigg(G(i\ep)\;\fract{\partial \Sigma(i\ep)}{\partial i\ep}\Bigg)
=\fint_{-\infty}^\infty \fract{d\ep}{2\pi}\;
\Tr\Bigg(G(i\ep)\;\fract{\partial}{\partial i\ep}\,\fract{\delta \Phi[G]}{\delta G(i\ep)}\Bigg)\\
& \equiv \fint_{-\infty}^\infty
\fract{d\ep}{2\pi}\;\fract{\partial}{\partial i\ep}\,
\Tr\Big(\ln X(i\ep)\Big)
\,,
\label{working hypothesis}
\eal
where, as we earlier mentioned, the regularizing exponential can be safely ignored in all expressions. Here the principal value matters: $\Tr\big(\ln X(i\ep)\big)$ jumps at $\ep=0$, so the $\int$ of its derivative vanishes, whereas the $\fint$ gives $I_L$. \\
Let us consider the functional
\beal
\Phi_L[X] &\coloneq \int_{-\infty}^\infty \fract{d\ep}{2\pi}\,
\Tr\Big(\ln X(i\ep)\Big)
= \int_{-\infty}^\infty \fract{d\ep}{2\pi}\,\bigg\{
\Tr\Big(\ln S(i\ep)\Big)-\Tr\Big(\ln G(i\ep)^{-1}\Big)\bigg\}\,,
\label{def F}
\eal
which we shall regard as a functional of $X(i\ep)$ that is in turn a functional of $G(i\ep)$. By the definition \eqn{def X},
$\Tr\big(\ln X(i\ep)\big)$ has the correct asymptotic behavior to make the integral in \eqn{def F}
absolutely convergent\footnote{We observe that for $|\ep|\to\infty$
\bealn
G(i\ep) \simeq \fract{1}{\;i\ep-H_\text{HF}\;}\simeq \fract{1}{i\ep}
+(i\ep)^{-2}\;H_\text{HF} + O(\ep^{-3})\;,
\eal
where $H_\text{HF}\equiv H_*(\ep\to\pm\infty)$ is the Hartree-Fock Hamiltonian.
Therefore,
\bealn
\Tr\Big(\ln S(i\ep)\Big) - \Tr\Big(\ln G(i\ep)^{-1}\Big) &\simeq
\sum_\ell\,\bigg(\ln i\ep - \fract{1}{i\ep}\,\big[H_\text{HF}\big]_{\ell\ell}\bigg)
-\Tr\Big(\ln\big(i\ep-H_\text{HF}\big)\Big)\simeq O(\ep^{-2})\,.
\eal
}, which guarantees the continuous translational invariance
\be
\int_{-\infty}^\infty \fract{d\ep}{2\pi}\,
\Tr\Big(\ln X(i\ep+i\omega)\Big) -
\int_{-\infty}^\infty \fract{d\ep}{2\pi}\,
\Tr\Big(\ln X(i\ep)\Big) \equiv 0\,.
\label{translational invariance}
\ee
To compute the functional derivative of $\Phi_L[X]$, we note that $\Tr\big(\ln X(i\ep)\big)=O(\ep^{-2})$ for $|\ep|\to\infty$ and is bounded for $\ep\to0^\pm$. Therefore, an integration by parts followed by Eqs.~\eqn{G diag 1} and \eqn{def X} yields
\beal
\Phi_L[X] &= -\int_{-\infty}^\infty \fract{d\ep}{2\pi}\;i\ep\;\fract{\partial}{\partial i\ep}\,\Tr\Big(\ln X(i\ep)\Big)\\
&= -\int_{-\infty}^\infty \fract{d\ep}{2\pi}\;i\ep\;\bigg\{\Tr\Big(G(i\ep)\Big)-\fract{\partial}{\partial i\ep}\,\Tr\Big(\ln G(i\ep)^{-1}\Big)\bigg\}\,,
\label{PhiL by parts}
\eal
which depends on $G$ only, $S$ having disappeared. 
We now consider a variation $\delta G(i\ep)$ that preserves the asymptotic behavior of $G(i\ep)$, and use $\delta\Tr\big(\ln G^{-1}\big)=-\Tr\big(G^{-1}\,\delta G\big)$. Integrating by parts the term that contains $\partial_{i\ep}\Tr\big(G^{-1}\delta G\big)$, we find
\bealn
\delta\Phi_L[X] &= -\int_{-\infty}^\infty \fract{d\ep}{2\pi}\;i\ep\;\bigg\{\Tr\Big(\delta G(i\ep)\Big)+\fract{\partial}{\partial i\ep}\,\Tr\Big(G(i\ep)^{-1}\,\delta G(i\ep)\Big)\bigg\}\\
&= \int_{-\infty}^\infty \fract{d\ep}{2\pi}\;\Tr\Big(\big(G(i\ep)^{-1}-i\ep\big)\,\delta G(i\ep)\Big)\,.
\eal
Therefore, through the Dyson equation \eqn{Dyson},
\be
\fract{\delta \Phi_L[X]}{\delta G(i\ep)} = G(i\ep)^{-1}-i\ep = -H_0-\Sigma(i\ep)\,,
\label{delta PhiL}
\ee
so that, at any $\ep\not=0$,
\be
\Tr\Bigg(G(i\ep)\;\fract{\partial}{\partial i\ep}\,\fract{\delta \Phi_L[X]}{\delta G(i\ep)}\Bigg)
= -\Tr\Bigg(G(i\ep)\;\fract{\partial}{\partial i\ep}\,\fract{\delta \Phi[G]}{\delta G(i\ep)}\Bigg)\,.
\label{chain rule PhiL}
\ee
We can therefore write
\be
\Phi[G] = \Phi'[G]-\Phi_L\big[X[G]\big]\,,
\label{Phi decomposition}
\ee
where $\Phi'[G]\coloneq\Phi[G]+\Phi_L\big[X[G]\big]$ has, by \eqn{self energy} and \eqn{delta PhiL}, the frequency-independent functional derivative $\delta\Phi'[G]/\delta G(i\ep)=-H_0$, so that
\be
\int_{-\infty}^\infty\fract{d\ep}{2\pi}\;\Tr\Bigg(G(i\ep)\;\fract{\partial}{\partial i\ep}\,\fract{\delta\Phi'[G]}{\delta G(i\ep)}\Bigg) \equiv 0\,.
\label{Phi prime}
\ee
Therefore, $\Phi_L[X]$ alone decides whether Luttinger's theorem holds or fails.

\subsection{Paradox of the functional chain rule and flaw in theorem's proof}
\label{Paradox of the functional chain rule and flaw in theorem's proof}

Since frequency translational invariance \eqn{translational invariance} holds,  
\bealn
\fract{\partial \Phi_L[X(i\ep+i\omega)]}{\partial i\omega} \equiv 0\,.
\eal
Computing the na\"{\i}ve functional derivative at $\omega=0$ leads to 
\beal
0 &\equiv \fract{\partial \Phi_L[X(i\ep+i\omega)]}{\partial i\omega}_{\big|\omega=0}
\overset{?}{=} \fint_{-\infty}^\infty 
\fract{d\ep}{2\pi}\;\fract{\partial}{\partial i\ep}\,
\Tr\Big(\ln X(i\ep)\Big) \equiv I_L\,,
\label{naive}
\eal
which is again Luttinger's proof of the theorem \cite{Luttinger}. However, 
\eqn{naive} contradicts the correct result \eqn{true IL}. The paradox is solved 
by noticing that $\Ima\,\Tr\big(\ln X(i\ep+i\omega)\big)$ has a jump $-2\pi Z$ at $\ep=-\omega$, which implies that 
\beal
\fract{\partial \Phi_L[X(i\ep+i\omega)]}{\partial i\omega} &= 
\fract{\partial}{\partial i\omega}\Bigg\{
\int_{-\infty}^{-\omega-0^+}\fract{d\ep}{2\pi}\,
\Tr\Big(\ln X(i\ep+i\omega)\Big)\\
& \qquad \qquad\qquad
+ \int_{-\omega+0^+}^{\infty}\fract{d\ep}{2\pi}\,
\Tr\Big(\ln X(i\ep+i\omega)\Big)\Bigg\}\\
&= \fract{1}{2\pi}\,\Big[\Ima\,\Tr\big(\ln X(i0^+)\big)
- \Ima\,\Tr\big(\ln X(-i0^+)\big)\Big]\\
&\qquad + \fint_{-\infty}^\infty 
\fract{d\ep}{2\pi}\;\fract{\partial}{\partial i\ep}\,
\Tr\Big(\ln X(i\ep+i\omega)\Big)\\
&\xrightarrow[\omega\to 0]{}\;  \fract{1}{2\pi}\,\Big[\Ima\,\Tr\big(\ln X(i0^+)\big)
- \Ima\,\Tr\big(\ln X(-i0^+)\big)\Big]
 + I_L
\equiv 0\,,
\eal
in perfect agreement with Eq.~\eqn{translational invariance}. This result is equivalent to assuming that  
\be
\fract{\partial \Phi_L[X(i\ep+i\omega)]}{\partial i\omega}_{\big|\omega=0}
\equiv \int_{-\infty}^\infty 
\fract{d\ep}{2\pi}\,\bigg\{
\Tr\Big[X(i\ep)^{-1}\,\partial_{i\ep} X(i\ep)\Big]_{\text{reg}}
-2\pi\,Z\,\delta(\ep)\bigg\}\,.
\label{correct functional derivative}
\ee
The regular piece integrates exactly to $Z$, which is flawlessly canceled by the $\delta$-function weight. This is the very element that is missing in the conventional proof of Luttinger's theorem \cite{Luttinger,Potthoff-2006}. As we demonstrated, the Luttinger integral is a pure boundary term, so the existence of the $\delta$-function anomaly cannot be ruled out a priori. In fact, it is essential to restore frequency translational invariance whenever $I_L\not=0$.

\subsection{The paradox in Matsubara formalism}
\label{The paradox in Matsubara formalism}
At $T\not=0$, $\Phi_L[X]$ reads
\bealn
\Phi_L[X(i\ep_n)] &= T\sum_n\,\Tr\Big(\ln X(i\ep_n)\Big)\,.
\eal
Translational invariance is trivially satisfied:
\bealn
\Phi_L[X(i\ep_n+i\omega_m)]-\Phi_L[X(i\ep_n)] &\equiv 0\,,& \omega_m &= 2m\,\pi T\,.
\eal
For $T\to 0$, thus $\omega_m\to 0$, it is tempting to make a Taylor expansion in  
\be
\Delta X(i\ep_n;i\omega_m) = X(i\ep_n+i\omega_m)-X(i\ep_n)\,.
\label{Delta X T not = 0}
\ee
Na\"{\i}vely, one would argue that $\Delta X\to 0$ for $T\to 0$ and conclude that  
\beal
0 &\equiv \lim_{T\to 0}\;\fract{\;\Phi_L[X(i\ep_n+i\omega_m)]-\Phi_L[X(i\ep_n)]\;}{i\omega_m}\\
&\overset{?}{\equiv} \lim_{T\to 0}\, \fract{T}{i\omega_m}\sum_n\, 
\sum_{k=1}^\infty\,\fract{\;(-1)^{k-1}\;}{k}\, 
\Tr\bigg\{ \Big(
X(i\ep_n)^{-1}\, \Delta X(i\ep_n;i\omega_m)\Big)^k\bigg\}\\
&\overset{?}{\equiv} \lim_{T\to 0}\, T\sum_n\,  
\Tr\left(
X(i\ep_n)^{-1}\, \fract{\;\Delta X(i\ep_n;i\omega_m)\;}{i\omega_m}\right)
\label{key of the proof}
\eal
which recovers only the regular part in \eqn{correct functional derivative} and thus violates 
frequency-translational invariance. There are several critical steps in \eqn{key of the proof}. First, the Taylor expansion in \eqn{key of the proof} derives formally from 
\bealn
\Tr\Big(\ln X(i\ep_n+i\omega_m)\Big) &= 
\Tr\bigg\{\ln\left[\Big(1+ \Delta X(i\ep_n;i\omega_m)\cdot X(i\ep_n)^{-1}\Big)\, 
X(i\ep_n)\right]\bigg\}\\
&= \Tr\bigg\{\ln\Big(1+ \Delta X(i\ep_n;i\omega_m)\, X(i\ep_n)^{-1}\Big)\bigg\}
+ \Tr\Big(\ln X(i\ep_n)\Big)\,,
\eal
which might not hold under the prescription of a fixed branch cut. In addition, the further 
expansion in powers of $\Delta X/X$ assumes that 
$\Delta X(i\ep_n;i\omega_m)\to 0$ for $T\to 0$, which may not occur when 
$\ep_n\,(\ep_n+\omega_m)<0$. \\
In reality, 
\beal
0 &\equiv \fract{\;\Phi_L[X(i\ep_n+i\omega_m)]-\Phi_L[X(i\ep_n)]\;}{i\omega_m}\\
&= \fract{T}{i\omega_m}\sum_n\, \bigg\{ \Tr\Big(\ln X(i\ep_n+i\omega_m)\Big)-
\Tr\Big(\ln X(i\ep_n)\Big)\bigg\}\\
&= \fract{T}{i\omega_m}\sum_n\, \theta\big(-\ep_n(\ep_n+\omega_m)\big)\,\bigg\{ \Tr\Big(\ln X(i\ep_n+i\omega_m)\Big)-
\Tr\Big(\ln X(i\ep_n)\Big)\bigg\}\\
&\qquad + \fract{T}{i\omega_m}\sum_n\, \theta\big(\ep_n(\ep_n+\omega_m)\big)\,\bigg\{ \Tr\Big(\ln X(i\ep_n+i\omega_m)\Big)-
\Tr\Big(\ln X(i\ep_n)\Big)\bigg\}\,.
\label{key of the proof 1}
\eal
The first term in \eqn{key of the proof 1} can be written, assuming $\omega_m>0$ and 
$\ep_n=(2n+1)\pi T$, as
\bealn
\fract{1}{2\pi i m}\,\sum_{n=0}^{m-1}\, 
\bigg\{ \Tr\Big(\ln X(i\ep_n)\Big)-
\Tr\Big(\ln X(-i\ep_n)\Big)\bigg\}\,,
\eal
which, for $T\to 0$ at fixed $m$, yields the anomaly in \eqn{correct functional derivative}. 
The second term in \eqn{key of the proof 1} can be safely expanded in $\Delta X$ and 
gives the regular contribution in \eqn{correct functional derivative}.

\subsection{Final remarks}
\label{Final remarks}
Now, we recall that $X$ is itself a functional of $G$; therefore, the fact that 
\bealn
I_L &= \fint_{-\infty}^\infty 
\fract{d\ep}{2\pi}\;\Tr\bigg(G(i\ep)\;\fract{\partial\Sigma(i\ep)}{\partial i\ep}\bigg)
= Z \not=0\,,
\eal
does not imply that the Luttinger-Ward functional loses its invariance under a shift of the internal frequency, which is merely a consequence of the particle number conservation, or that the skeleton-diagram formalism \cite{footnote} breaks down when correlations are strong. It only means 
that the formal Taylor expansion is incorrect unless one adds explicitly an anomaly term, i.e., a $\delta$-function in $\ep$, see \eqn{correct functional derivative}, which enforces the invariance.\\
By Eq.~\eqn{pointwise}, $I_L$ is contributed by $\partial_\ep\,\Ima\,\Tr\big(\ln X(i\ep)\big)$, whose regular part may behave as the difference between two Lorentzian functions. Consider, as an illustrative example, the function
\bealn
F(\ep) &= \fract{\gamma_1}{\;\ep^2+\gamma_1^2\;}-
\fract{\gamma_2}{\;\ep^2+\gamma_2^2\;}\;,
\eal
whose integral vanishes only if $\gamma_1\gamma_2>0$, corresponding to the regime of validity 
of Luttinger's theorem. Once $\gamma_1\gamma_2<0$, Luttinger's theorem is violated even though $F(\ep)$ does not exhibit any singularity. In other words, it is impossible to detect any indication of Luttinger's theorem violation from the local behavior of $F(\ep)$ or, equivalently, of $G(i\ep)\,\partial_{i\ep}\Sigma(i\ep)$, except right at the point where the theorem ceases to be valid, which is when either $\gamma_1$ or $\gamma_2$ vanishes.  \\
We also remark that, while the trace $\Tr\big(G(i\ep)\,\partial_{i\ep}\Sigma(i\ep)\big)$ is exactly the derivative of a logarithm, see \eqn{pointwise}, the individual diagonal elements of $G(i\ep)\,\partial_{i\ep}\Sigma(i\ep)$ generally are not: several of them conspire to reconstruct it. \\

\noindent
We conclude by briefly discussing Refs.~\cite{Yunoki-PRB2017} and \cite{mio-3}, which also claim that $I_L$ is a boundary term. The authors of Ref.~\cite{Yunoki-PRB2017} assume that 
$I_L=N-N_L$ is equivalent to $N_0 - N_L$, where $N_0$ is the electron number in the non-interacting ground state. However, the assumption $N=N_0$ is generally not true at a chemical potential common to the interacting and non-interacting systems.  
By contrast, in an earlier work by one of the authors \cite{mio-3} it was concluded that $I_L$ is a boundary term from the conjecture that the limit $T\to 0$ in \eqn{LT invariance T=0} is singular. Specifically, Ref.~\cite{mio-3} takes for granted that Eq.~\eqn{LT invariance} implies 
\beal
0& \equiv \lim_{T\to 0}\,\fract{\;\Phi\big[G(i\ep_n) +\Delta G(i\ep_n;i\omega_m)\big] -\Phi\big[G(i\ep_n)\big]\;}{i\omega_m} \\
& \overset{?}{\equiv} \lim_{T\to 0}\,
T\,\sum_n\, \Tr\bigg(\Sigma(i\ep_n)\;\fract{\;\Delta G(i\ep_n;i\omega_m)\;}{i\omega_m}\bigg) 
\,,
\label{Jan conjecture}
\eal
i.e. that the functional chain rule applies at $T\not=0$, but argues that the limit $T\to 0$ does not correspond to $-I_L$. 
As we discussed in Sec.~\ref{The paradox in Matsubara formalism},  this is only one piece of the puzzle, specifically the fact that $\Delta X$ in \eqn{Delta X T not = 0} might not vanish for $T\to 0$, thus invalidating the Taylor expansion. The other piece of the puzzle is that the functional chain rule by itself may miss an anomaly that is crucial to enforce frequency-translation invariance. We also mention that 
a pathology in the $T\to 0$ limit was originally proposed by Potthoff \cite{Potthoff-2006} as an explanation of Luttinger's theorem violation. \\
Our proof that $I_L$ is a boundary term is instead exact and does not rely on any \textit{a priori} assumption .

\section{Luttinger's theorem and the Streda formula}
\label{Sec3}
In this section we elaborate on the interesting results by Gavensky, Sachdev and Goldman \cite{PhysRevLett.131.236601} and extend them further to elucidate what the violation 
of Luttinger's theorem actually implies in a topological correlated insulator. \\
Since translational symmetry is important, we assume here, without loss of generality, a basis of $M$ topologically trivial Bloch wavefunctions $\phi_{\ell\bk}(\br)$ ($\ell=1,\dots,M$). Therefore, the 
non-trivial topology is entirely encoded in the Green's function $G(i\ep,\bk)$, an $M\times M$ 
complex matrix for each momentum $\bk$, and in the functional derivatives of the LW functional.
In addition, we consider a two-dimensional insulator.  \\
An important role hereafter will be played by the irreducible vertex in the particle-hole channel with transferred imaginary frequency $i\omega$ and momentum $\bq$. This is defined through 
\beal
&\Gamma_0\big(i\ep+i\omega,\bk+\bq,\ell;i\ep',\bkp,\ell'|i\ep'+i\omega,\bkp+\bq,m';
i\ep,\bk,m\big) \\
&\qquad \coloneq \fract{\delta^2\Phi[G]}
{\delta G_{m\ell}(i\ep,\bk;i\ep+i\omega,\bk+\bq)\,
\delta G_{m'\ell'}(i\ep'+i\omega,\bkp+\bq;i\ep',\bkp)} 
\;,
\eal
where $\ell$ and $\ell'$ identify the incoming legs, while $m$ and $m'$ the outgoing ones.  
The vertex satisfies the following identities:
\beal
&\Gamma_0\big(i\ep+i\omega,\bk+\bq,\ell;i\ep',\bkp,\ell'|i\ep'+i\omega,\bkp+\bq,m';
i\ep,\bk,m\big)\\
&\qquad = \Gamma_0\big(i\ep',\bkp,\ell';i\ep+i\omega,\bk+\bq,\ell|
i\ep,\bk,m;i\ep'+i\omega,\bkp+\bq,m'\big)\,,\\
&\Gamma_0\big(i\ep+i\omega,\bk+\bq,\ell;i\ep',\bkp,\ell'|i\ep'+i\omega,\bkp+\bq,m';
i\ep,\bk,m\big)^*\\
&\qquad = \Gamma_0\big(-i\ep,\bk,m;-i\ep'-i\omega,\bkp+\bq,m'|-i\ep',\bkp,\ell';
-i\ep-i\omega,\bk+\bq,\ell\big)\,.
\label{vertex properties 1}
\eal
Moreover, if the system is invariant under time-reversal symmetry $\mathcal{T}$, such that 
$\mathcal{T}\big(\phi_{\ell\bk}(\br)\big) = \phi_{\bar{\ell}-\bk}(\br)$, then 
\beal
&\Gamma_0\big(i\ep+i\omega,\bk+\bq,\ell;i\ep',\bkp,\ell'|i\ep'+i\omega,\bkp+\bq,m';
i\ep,\bk,m\big)\\
&\qquad \equiv \Gamma_0\big(-i\ep-i\omega,-\bk-\bq,\bar{\ell};-i\ep',-\bkp,\bar{\ell}\,'|-i\ep'-i\omega,-\bkp-\bq,\bar{m}';
-i\ep,-\bk,\bar{m}\big)^*\,,
\label{vertex properties TR}
\eal
as well as
\beal
G_{nm}(i\ep,\bk) \equiv G_{\bar{n}\bar{m}}(-i\ep,-\bk)^*\,.
\label{property G TR}      
\eal
To simplify the notation, we find it convenient to introduce two four-index tensors, 
$\Gamma^S_{0}(i\ep,\bk;i\ep',\bkp|i\omega,\bq)$ and $\Gamma^A_{0}(i\ep,\bk;i\ep',\bkp|i\omega,\bq)$
with the following tensor elements 
\beal
&\Gamma^S_{0|\ell m,\ell' m'}(i\ep,\bk;i\ep',\bkp|i\omega,\bq) \\
&\qquad \coloneq \fract{1}{2}\,\Big(
\Gamma_0(i\ep+i\omega,\bk+\bq,\ell;i\ep',\bkp,\ell'|i\ep'+i\omega,\bkp+\bq,m';i\ep,\bk,m)\\
&\qquad\qquad\qquad  + 
\Gamma_0(i\ep'+i\omega,\bkp+\bq,\ell';i\ep,\bk,\ell|i\ep+i\omega,\bk+\bq,m;i\ep',\bkp,m')\Big)\,,\\
&\Gamma^A_{0|\ell m,\ell'm'}(i\ep,\bk;i\ep',\bkp|i\omega,\bq) \\
&\qquad \coloneq \fract{1}{2}\,\Big(
\Gamma_0(i\ep+i\omega,\bk+\bq,\ell;i\ep',\bkp,\ell'|i\ep'+i\omega,\bkp+\bq,m';i\ep,\bk,m)\\
&\qquad\qquad\qquad  - 
\Gamma_0(i\ep'+i\omega,\bkp+\bq,\ell';i\ep,\bk,\ell|i\ep+i\omega,\bk+\bq,m;i\ep',\bkp,m')\Big)\,,
\label{Gamma S and A}
\eal
where $S$ and $A$ stand for symmetric and antisymmetric properties under
\bealn
(i\ep,\bk) \leftrightarrow (i\ep',\bkp)\;\land\; (\ell,m) \leftrightarrow (\ell',m')\,.
\eal 
In addition, we indicate the matrix product by $\cdot$ and the tensor contraction by $\tdot$, 
meaning that 
\bealn
\Gamma_0\tdot Y = \sum_{\ell'm'}\,
\Gamma_{0|\ell m,\ell'm'}\,Y_{m' \ell'}
\eal
is a matrix. Moreover, whenever we need to denote symbolically the tensor contraction and the integral over internal frequencies and momenta, we use the symbol $\odot$. We also adopt the following conventions:
\bealn
\Gamma^{S/A}_0(i\ep,\bk;i\ep',\bkp|0,0) &\coloneq 
\Gamma^{S/A}_0(i\ep,\bk;i\ep',\bkp)\,,\\
\partial_{i\omega} \Gamma^{S/A}_0(i\ep,\bk;i\ep',\bkp|i\omega,\bq)_{\big|\omega=\bq=0} 
&\coloneq \dot{\Gamma}^{S/A}_0(i\ep,\bk;i\ep',\bkp)\,,\\
\partial_{q_\mu} \Gamma^{S/A}_0(i\ep,\bk;i\ep',\bkp|i\omega,\bq)_{\big|\omega=\bq=0} 
&\coloneq \Gamma^{S/A}_{0\mu}(i\ep,\bk;i\ep',\bkp)\,.
\eal
We emphasize that, although 
$\Gamma^A_0(i\ep,\bk;i\ep',\bkp)$ trivially vanishes, its derivatives 
$\dot{\Gamma}^A_0(i\ep,\bk;i\ep',\bkp)$ and $\Gamma^A_{0\mu}(i\ep,\bk;i\ep',\bkp)$ have no reason to vanish either.\\

\noindent 
With the above notation, the representation within the skeleton-diagram formalism of the Hall conductance in two dimensions reads \cite{Andrea-PRB2023}
\beal
\sigma_\text{H} &= \sigma_\text{H\,L}
+\fract{i e^2}{2V^2}\sum_{\bk\bkp}\,\int\fract{d\ep}{2\pi}\,\fract{d\ep'}{2\pi}\,\ep_{\mu\nu}\,
\Tr\Big(\partial_\mu G(i\ep,\bk)\cdot \dot{\Gamma}^A_{0}(i\ep,\bk;i\ep',\bkp)\tdot
\partial_\nu G(i\ep',\bkp)\Big)\\
&\equiv -\fract{ec}{V}\;\fract{\partial N_L}{\partial B} - \fract{ec}{V}\;\fract{\partial I_L}{\partial B} = \fract{e^2}{2\pi}\;\nu_{\text{H}\,L} + \fract{e^2}{2\pi}\;\nu^I_{\text{H}\,L}\;,
\label{Hall-1}
\eal 
where $\nu_\text{H\,L}$ has been defined in \eqn{I and M 1} and $\nu^I_{\text{H}\,L}$ 
quantifies the violation of Luttinger's theorem within the Streda formula \cite{PhysRevLett.131.236601}. One can readily verify 
that $\sigma_\text{H}$ indeed vanishes if the system is time-reversal invariant, namely when Eqs.~\eqn{property G TR} and \eqn{vertex properties TR} are satisfied. 
The equivalence 
between $\nu_{\text{H}\,L}$ and the Ishikawa-Matsuyama-Volovik topological invariant has been discussed in \cite{Andrea-PRB2023} and explicitly proven in \cite{PhysRevLett.131.236601}.
In what follows, we demonstrate that $-ec\,\partial_B I_L/V$ yields exactly the integral in \eqn{Hall-1}. Although this relation may appear straightforward, its explicit verification is highly instructive, as it confirms that the skeleton-diagram formalism remains fully consistent with the Streda formula, a strict thermodynamic identity.

\subsection{The Luttinger integral in a magnetic field}
\label{The Luttinger integral in a magnetic field}

Let us now consider a correlated two-dimensional insulator in a weak perpendicular magnetic field $B$. Following \cite{PhysRevLett.131.236601}, and upon introducing the Peierls phase factor
\beal
\phi(\bR_1,\bR_2) &= -\fract{e}{c}\,\int_{\bR_2}^{\bR_1} d\br\cdot\bd{A}(\br)\,,
\eal
the non-interacting Hamiltonian can be written as 
\bealn
H(\bR_1,\bR_2) &= \esp{i\phi(\bR_1,\bR_2)}\;H_0(\bR_1,\bR_2)\,,
\eal
where $H_0(\bR_1,\bR_2)$ is its expression in absence of $B$ and $\bR_i$ are lattice sites. Correspondingly, one can write \cite{PhysRevLett.131.236601} 
\bealn
G(i\ep;\bR_1,\bR_2) &= \esp{i\phi(\bR_1,\bR_2)}\;G_B(i\ep;\bR_1,\bR_2)\,,&
\Sigma(i\ep;\bR_1,\bR_2) &= \esp{i\phi(\bR_1,\bR_2)}\;\Sigma_B(i\ep;\bR_1,\bR_2)\,,
\eal
where the functions with the $B$ subscript are gauge invariant, as well as  
\beal
G_B(i\ep;\bR_1,\bR_2) &= G^{(0)}(i\ep;\bR_1,\bR_2) + B\,G^{(1)}(i\ep;\bR_1,\bR_2)\,,
\label{GB0 and GB1}
\eal
and similarly for the self-energy. The first term on the right hand side of \eqn{GB0 and GB1} is the Green's function at zero field while the second is its first order correction. It was found in \cite{PhysRevLett.131.236601} that 
the Fourier transform of $G^{(1)}$ satisfies
\beal
G^{(1)}(i\ep,\bk) &= i\,\fract{e}{2c}\,\ep_{\mu\nu}\, G^{(0)}(i\ep,\bk)\cdot
\partial_\mu G^{(0)}(i\ep,\bk)^{-1}\cdot \partial_\nu G^{(0)}(i\ep,\bk)\\
&\qquad + G^{(0)}(i\ep,\bk)\cdot\Sigma^{(1)}(i\ep,\bk)\cdot G^{(0)}(i\ep,\bk)\,.
\label{Lucilla 1}
\eal

\subsection{Magnetic field corrections to the self-energy}
\label{Magnetic field corrections to the self-energy}

Now, we extend the results of \cite{PhysRevLett.131.236601}.
The Luttinger-Ward functional $\Phi[G]$ does not depend explicitly on $B$, which enters only through $G$, so that
$\Sigma[G]=\delta\Phi/\delta G$ is the same functional at any field. 
In short notation, 
\be
\Sigma_B = \esp{-i\phi(B)}\;\Sigma\big[\esp{i\phi(B)}\;G_B\big]\,,
\label{NA SigmaB}
\ee
so that, if we write $\phi(B)=B\,\phi^{(1)}$, ignore 
for the moment the frequency, and expand \eqn{NA SigmaB} to first
order in $B$, the chain rule yields  $\Sigma^{(1)}=\Sigma^{(1)}_\text{direct}+\Sigma^{(1)}_\text{exchange}$, where $\Sigma^{(1)}_\text{direct}=\Gamma_0\odot G^{(1)}$, while the contribution of the Peierls phases reads
\beal
\Sigma^{(1)}_\text{exchange}(\bR_1,\bR_2) &= i\,\sum_{\bR_3,\bR_4}\,\Gamma_0(\bR_1,\bR_2;\bR_3,\bR_4)\tdot\phi^{(1)}(\bR_4,\bR_3)\,G^{(0)}(\bR_4,\bR_3)\\
&\qquad -i\,\phi^{(1)}(\bR_1,\bR_2)\,\Sigma^{(0)}(\bR_1,\bR_2)\,.
\label{NA chain rule}
\eal
Equation \eqn{NA chain rule} can be evaluated in any gauge, because the Luttinger-Ward functional is gauge invariant\footnote{Indeed, gauge invariance implies that 
\bealn
\big(\bR_1-\bR_2\big)\,\Sigma^{(0)}(\bR_1,\bR_2) 
= \sum_{\bR_3\bR_4}\, \Gamma_0(\bR_1,\bR_2;\bR_3,\bR_4)\tdot 
\big(\bR_4-\bR_3\big)\,G^{(0)}(\bR_4,\bR_3)\,,
\eal
which can be employed to demonstrate that, whatever gauge choice we make, we always end up in Eq.~\eqn{NA exchange R}.
}.
In particular, in the symmetric gauge, $\bd{A}(\br)=\fract12\,\bd{B}\times(\br-\br_0)$, 
\begin{equation*}
\phi^{(1)}(\bR_a,\bR_b)=\fract{e}{2c}\,(\bR_a-\br_0)\times(\bR_b-\br_0)\,,
\end{equation*}
and vanishes when $\br_0$ lies on the line through $\bR_a$ and $\bR_b$. We therefore center the gauge at the midpoint of the external pair,
$\br_0=(\bR_1+\bR_2)/2$, so that $\phi^{(1)}(\bR_1,\bR_2)=0$, only the first term in \eqn{NA chain rule} survives and reads 
\beal
\Sigma^{(1)}_\text{exchange}(\bR_1,\bR_2) &= i\,\fract{e}{4c}\,\sum_{\bR_3,\bR_4}\,\big(\bR_3+\bR_4-\bR_1-\bR_2\big)\times \big(\bR_3-\bR_4\big)\\
&\qquad\qquad \qquad \qquad \quad \Gamma_0(\bR_1,\bR_2;\bR_3,\bR_4)\tdot G^{(0)}(\bR_4,\bR_3)\,.
\label{NA exchange R}
\eal
We can now take the Fourier transform and, after some lengthy but straightforward calculations, obtain 
\be
\Sigma^{(1)}_\text{exchange}(\bk) = 
- \fract{ie}{2c}\,\fract{1}{V}\,\sum_{\bkp}\,\ep_{\mu\nu}\,\Gamma_{0\mu}^A(\bk;\bkp)\tdot\partial_{\nu} G^{(0)}(\bkp)\,.
\label{new idea 5 before}
\ee
Collecting all results, we thus find that
\be
\Sigma^{(1)}(\bk) = \fract{1}{V}\,\sum_{\bkp}\,\Gamma_0(\bk;\bkp)\tdot G^{(1)}(\bkp)
- \fract{ie}{2c}\,\fract{1}{V}\,\sum_{\bkp}\,\ep_{\mu\nu}\,\Gamma_{0\mu}^A(\bk;\bkp)\tdot\partial_{\nu} G^{(0)}(\bkp)\,.
\label{new idea 5}
\ee
We rewrite Eq.~\eqn{Lucilla 1} as 
\beal
G^{(1)}(\bk) &= i\,\fract{e}{2c}\,\ep_{\mu\nu}\, G^{(0)}(\bk)\cdot
\partial_\mu G^{(0)}(\bk)^{-1}\cdot \partial_\nu G^{(0)}(\bk)
 + G^{(0)}(\bk)\cdot\Sigma^{(1)}(\bk)\cdot G^{(0)}(\bk)\\
&\coloneq \Delta G(\bk) + G^{(0)}(\bk)\cdot\Sigma^{(1)}(\bk)\cdot G^{(0)}(\bk)\,,
\label{Lucilla 2}
\eal
where $\Delta G(\bk)$ looks like the simple variation of $G$ in presence of $B$.
It follows that, setting $G^{(0)}\equiv G$,  
\bealn
\Sigma^{(1)}(\bk) &= \Sigma^{(1)}_\text{direct}(\bk)
+\Sigma^{(1)}_\text{exchange}(\bk)\\
& =\fract{1}{V}\,\sum_\bkp\, 
\Gamma_0(\bk;\bkp)\tdot G^{(1)}(\bkp)
- \fract{ie}{2c}\,\fract{1}{V}\,
\sum_{\bkp} \,  \ep_{\mu\nu}\,\Gamma_{0\mu}^A(\bk;\bkp) \tdot 
 \partial_{\nu} G(\bkp)\\
 & = \fract{1}{V}\,\sum_\bkp\, 
\Gamma_0(\bk;\bkp)\tdot \Delta G(\bkp)+ \fract{1}{V}\,\sum_\bkp\, 
\Gamma_0(\bk;\bkp)\tdot G(\bkp)\cdot\Sigma^{(1)}(\bkp)\cdot G(\bkp)\\
&\qquad\qquad\qquad  +\Sigma^{(1)}_\text{exchange}(\bk)
\,,
\eal
which we shortly write as
\bealn
\Big(\mathbb{I}- \Gamma_0\odot R\Big)\odot \Sigma^{(1)}
&= \Gamma_0\odot\Delta G +\Sigma^{(1)}_\text{exchange}\,,
\eal
where $R=G\otimes G$ is a tensor.
Since, if $\Gamma$ is the reducible vertex, the following equations hold: 
\bealn
&\Big(\mathbb{I}- \Gamma_0\odot R\Big)^{-1}\odot \Gamma_0 \equiv \Gamma\,,\\
&\Big(\mathbb{I}- \Gamma_0\odot R\Big)^{-1} - \mathbb{I} =
\Big(\mathbb{I}- \Gamma_0\odot R\Big)^{-1}\odot\bigg\{
\mathbb{I} -\Big(\mathbb{I}- \Gamma_0\odot R\Big)\bigg\}
= \Gamma\odot R\,,
\eal
we find that 
\bealn
\Sigma^{(1)}&=\Sigma^{(1)}_\text{direct}
+\Sigma^{(1)}_\text{exchange}
= \Gamma\odot\Delta G +\Big(\mathbb{I}- \Gamma_0\odot R\Big)^{-1}\odot\Sigma^{(1)}_\text{exchange}\,,
\eal
thus 
\bealn
\Sigma^{(1)}_\text{direct} &= \Gamma\odot\Delta G + \Gamma\odot R\odot \Sigma^{(1)}_\text{exchange}\,.
\label{Lucilla 3}
\eal
On the other hand, 
\bealn
G^{(1)} &= \Delta G + G\cdot\Sigma^{(1)}_\text{direct}\cdot G + 
G\cdot \Sigma^{(1)}_\text{exchange}\cdot G\\
&= \Delta G + G\cdot\Gamma\odot\Delta G\cdot G
+ G\cdot \Gamma\odot R\odot \Sigma^{(1)}_\text{exchange}\cdot G
+ 
G\cdot \Sigma^{(1)}_\text{exchange}\cdot G\\
&\equiv \Delta G + G\cdot \Sigma^{(1)}_\text{exchange} \cdot G
 + G\cdot \Gamma\odot\Big(\Delta G + G\cdot \Sigma^{(1)}_\text{exchange} \cdot G\Big)
\cdot G\,.
\eal
This looks like an equation for a three-leg vertex with the \textit{bare} value
\beal
\Lambda_B^0(i\ep,\bk) &\coloneq -i\pi\;\ep_{\mu\nu}\,\Bigg\{
\partial_\mu G(i\ep,\bk)^{-1}
\cdot G(i\ep,\bk) \cdot \partial_\nu G(i\ep,\bk)^{-1}\\
&\qquad\qquad\qquad   +\fract{1}{V}\,\sum_\bkp\,\int \fract{d\epsilon'}{2\pi}\;
\Gamma^A_{0\mu}(i\ep,\bk;i\ep',\bkp)\tdot \partial_\nu G(i\ep',\bkp)\Bigg\}\,,
\label{Lucilla 4}
\eal
the fully interacting one
\be
\Lambda_B(i\ep,\bk) = \Lambda_B^0(i\ep,\bk) 
+\fract{1}{V}\,\sum_\bkp\,\int \fract{d\epsilon'}{2\pi}\;
\Gamma(i\ep,\bk;i\ep',\bkp)\tdot G(i\ep',\bkp) \cdot \Lambda_B^0(i\ep',\bkp)\cdot G(i\ep',\bkp)\,,
\label{Lucilla 5}
\ee
so that 
\be
G^{(1)}(i\ep,\bk) = \fract{e}{2\pi c}\,G(i\ep,\bk)\cdot \Lambda_B(i\ep,\bk) \cdot G(i\ep,\bk)\,.
\label{Lucilla 6}
\ee
If we assume, for simplicity, that $\Tr(\dots)$ includes the sum over $\bk$ and the integral over $\ep$ times the regularizing factor, 
and note that 
\bealn
1 &= \partial_{i\ep} G(i\ep,\bk)^{-1} + \partial_{i\ep} \Sigma(i\ep,\bk) 
\coloneq \dot{G}^{-1} + \dot{\Sigma}\,,
\eal
which is precisely the trick used to derive Luttinger's theorem, then, 
using Eq.~\eqn{simple WT identity}, 
\beal
\Tr\Big(G^{(1)}\Big) &= \fract{e}{2\pi c}\,\Tr\Big(G\cdot G\cdot \Lambda_B\Big)\\
&= \fract{e}{2\pi c}\,\Tr\Big(G\cdot \dot{G}^{-1} \cdot G\cdot \Lambda_B\Big)
+ \fract{e}{2\pi c}\,\Tr\Big(G\cdot \dot{\Sigma} \cdot G\cdot \Lambda_B\Big)\\
&= -\fract{e}{2\pi c}\,\Tr\Big(\dot{G}\cdot \Lambda_B\Big) 
+\fract{e}{2\pi c}\, \Tr\bigg(G\cdot \Big( \Gamma_0 \odot \dot{G}\Big)\cdot G\cdot \Lambda_B\bigg)\\
&= -\fract{e}{2\pi c}\,\Tr\Big(\dot{G}\cdot \Lambda^0_B \Big)
-\fract{e}{2\pi c}\,\Tr\Big(\dot{G}\cdot \Gamma \odot R \odot \Lambda^0_B\Big)\\
&\qquad +\fract{e}{2\pi c}\,\Tr\Big(\dot{G}\cdot \Gamma_0 \odot R \odot \Lambda^0_B\Big)
+ \fract{e}{2\pi c}\,\Tr\Big(\dot{G}\cdot \Gamma_0 \odot R \odot \Gamma\odot R\odot \Lambda^0_B\Big)\\
&= -\fract{e}{2\pi c}\,\Tr\Big(\dot{G}\cdot \Lambda^0_B\Big)\,,
\label{Lucilla 7}
\eal
since $-\Gamma\odot R +\Gamma_0\odot R + \Gamma_0\odot R\odot \Gamma\odot R = 0$.
More explicitly, and upon defining $N(B) \coloneq N^{(0)} + B\,N^{(1)}$, 
$N_L(B) \coloneq N_L^{(0)} + B\,N_L^{(1)}$ and $I_L(B) \coloneq I_L^{(0)} + B\,I_L^{(1)}$, 
\bealn
\Tr\Big(G^{(1)}\Big) &= N^{(1)} = -\fract{e}{2\pi c}\,\Tr\Big(\dot{G}\cdot \Lambda^0_B\Big)\\
&= \fract{ie}{2c}\, \sum_\bk\, \int\fract{d\ep}{2\pi}\;\ep_{\mu\nu}\,
\Tr\Big(\partial_{i\ep} G(i\ep,\bk)\cdot \partial_\mu G(i\ep,\bk)^{-1}
\cdot G(i\ep,\bk) \cdot \partial_\nu G(i\ep,\bk)^{-1}\Big)\\
&\qquad +\fract{ie}{2c}\,\fract{1}{V}\,\sum_{\bk\bkp}\, \int\fract{d\ep}{2\pi}\,\fract{d\ep'}{2\pi}\,\ep_{\mu\nu}\,
\Tr\Big(\partial_{i\ep} G(i\ep,\bk) \cdot \Gamma^A_{0\mu}(i\ep,\bk;i\ep',\bkp)\tdot  \partial_\nu G(i\ep',\bkp)\Big)\\
&= N_L^{(1)} +\fract{ie}{2c}\,\fract{1}{V}\,\sum_{\bk\bkp}\, \int\fract{d\ep}{2\pi}\,\fract{d\ep'}{2\pi}\,\ep_{\mu\nu}\,
\Tr\Big(\partial_{i\ep} G(i\ep,\bk) \cdot \Gamma^A_{0\mu}(i\ep,\bk;i\ep',\bkp)\tdot  \partial_\nu G(i\ep',\bkp)\Big)\\
&\equiv N_L^{(1)} + I_L^{(1)}\,,
\eal
where one readily demonstrates that 
\beal
I_L^{(1)} &\equiv \Tr\Big(G^{(1)}\cdot\dot{\Sigma}^{(0)}\Big) 
+ \Tr\Big(G^{(0)}\cdot\dot{\Sigma}^{(1)}\Big)\,.
\label{formal IL(1)}
\eal
Using the Ward-Takahashi identity derived in Appendix \ref{Ward-Takahashi identity}, see Eq.~\eqn{non trivial WT identity}, 
we find that  
\beal
&\fract{1}{V}\,\sum_{\bk \bkp}\,\int\fract{d\ep}{2\pi}\,\fract{d\ep'}{2\pi}\;\ep_{\mu\nu}\,
\Tr\Big(\partial_\mu G(i\ep,\bk)\cdot\dot{\Gamma}^A_0(i\ep,\bk;i\ep',\bkp)\tdot\partial_\nu G(i\ep',\bkp)\Big)\\
&\qquad =-\fract{1}{V}\,\sum_{\bk\bkp}\,\int \fract{d\ep}{2\pi}\,\fract{d\ep'}{2\pi}\,
\ep_{\mu\nu}\,
\Tr\Big(\partial_{i\ep}G(i\ep,\bk)\cdot 
\Gamma_{0\mu}^A(i\ep,\bk;i\ep',\bkp) \tdot 
 \partial_{\nu} G(i\ep',\bkp)\Big)\,,
\label{Important 4 bis}
\eal
and thereby conclude that   
\beal
-I_L^{(1)} &= -\fract{ie}{2c}\,\fract{1}{V}\,\sum_{\bk\bkp}\, \int\fract{d\ep}{2\pi}\,\fract{d\ep'}{2\pi}\,\ep_{\mu\nu}\,
\Tr\Big(\partial_{i\ep} G(i\ep,\bk) \cdot \Gamma^A_{0\mu}(i\ep,\bk;i\ep',\bkp)\tdot  \partial_\nu G(i\ep',\bkp)\Big)\\
&= \fract{ie}{2c}\,\fract{1}{V}\,\sum_{\bk\bkp}\,\int\fract{d\ep}{2\pi}\,\fract{d\ep'}{2\pi}\;\ep_{\mu\nu}\,
\Tr\Big(\partial_\mu G(i\ep,\bk)\cdot\dot{\Gamma}^A_0(i\ep,\bk;i\ep',\bkp)\tdot\partial_\nu G(i\ep',\bkp)\Big)\,,
\label{IL(1)}
\eal
which is precisely the correction to the Ishikawa-Matsuyama-Volovik expression for the Hall conductance in \eqn{Hall-1}. \\

\noindent
We introduce a three-component index vector $\mu,\nu,\rho\in\{0,1,2\}$, where 
$\partial_0 \coloneq \partial_{i\ep}$, $\partial_1\coloneq\partial_{k_x}$, and $\partial_2\coloneq\partial_{k_y}$, 
as well as    
\bealn
\gamma^A_{0\,\mu=0}(i\ep,\bk;i\ep',\bkp) &\coloneq \dot{\Gamma}^A_0(i\ep,\bk;i\ep',\bkp)\,,\\
\gamma^A_{0\,\mu=1,2}(i\ep,\bk;i\ep',\bkp) &\coloneq \Gamma^A_{0\,\mu=1,2}(i\ep,\bk;i\ep',\bkp)\,, 
\eal
which enables rewriting the first-order magnetic correction \eqn{Hall-1} compactly as 
\beal
\nu^I_{H\,L} &= -i\,\fract{\pi}{3}\,\fract{1}{V^2}\,\sum_{\bk\bkp}\, \int\fract{d\ep}{2\pi}\,\fract{d\ep'}{2\pi}\,\ep_{\mu\nu\rho}\,
\Tr\Big(\partial_{\mu} G(i\ep,\bk) \cdot \gamma^A_{0\nu}(i\ep,\bk;i\ep',\bkp)\tdot  \partial_\rho G(i\ep',\bkp)\Big)\,,
\label{IL(1) dream}
\eal
where $\ep_{\mu\nu\rho}$ is the completely antisymmetric Levi-Civita tensor in three dimensions.
Equation \eqn{IL(1) dream} represents the exact interaction-induced correction to the 
Ishikawa-Matsuyama-Volovik invariant $\nu_{H\,L}$ when strong 
correlations in a topological insulator drive a violation of Luttinger's theorem. Unlike $\nu_{H\,L}$, 
which is quantized in integer values, the interaction correction $\nu^I_{H\,L}$ can also acquire proper fractional values when the actual Hall number is fractional. Not surprisingly, 
$\nu^I_{H\,L}$ is invariant under any variation of the Green's function that leaves the system 
an insulator without Luttinger surfaces, as we demonstrate in Appendix \ref{Invariance property}.

\section{The Luttinger Integral as a novel topological invariant} 
\label{The Luttinger Integral as a novel topological invariant}

The inherent consistency between the skeleton-diagram formalism and the 
thermodynamic Streda formula raises a fundamental question: what precise 
mathematical conditions dictate the vanishing or non-vanishing of $I_L$? \\ 

\noindent
To this end, we here consider explicitly both metals and insulators under the assumption that \eqn{main assumption} holds. We decompose the single-particle Green's function and self-energy as in Eqs.~\eqn{LT: G1 and G2} and \eqn{LT: Sigma1 and Sigma2},
\beal
G(i\ep) &= G_1(\ep)+i\,G_2(\ep)\,,& \Sigma(i\ep) &= \Sigma_1(\ep)+i\,\Sigma_2(\ep)\,,
\label{G1 G2}
\eal
where we recall that $G_1(\ep)$ and $\Sigma_1(\ep)$ are hermitian and even in $\ep$, while $G_2(\ep)$ and $\Sigma_2(\ep)$ are also hermitian but odd in $\ep$ and negative for $\ep>0$. 
We also recall that in periodic systems the trace includes momentum summation as well as the sum over spin and Bloch waves.  \\

\noindent
We begin by investigating under which conditions $I_L$ is invariant upon a smooth variation $\delta G(i\ep)$ of $G(i\ep)$ that vanishes at $\ep=\pm\infty$. If $G(i\ep)\to G(i\ep)+\delta G(i\ep)$, 
$I_L\to I_L + \delta I_L$ where 
\be
\delta I_L = \fint_{-\infty}^\infty \fract{d\ep}{2\pi}\,
\Tr\Big(\delta G(i\ep)\cdot\partial_{i\ep}\Sigma(i\ep)\Big) 
+\fint_{-\infty}^\infty \fract{d\ep}{2\pi}\,
\Tr\Big(G(i\ep)\cdot \partial_{i\ep}  \delta\Sigma(i\ep)\Big)
\coloneq \delta I_L^{(1)} + \delta I_L^{(2)}\,.
\label{first step}
\ee
By means of Eq.~\eqn{simple WT identity}, we can write
\beal
\delta I_L^{(1)} &= \fint_{-\infty}^\infty \fract{d\ep}{2\pi}\,
\Tr\Big(\delta G(i\ep)\cdot\partial_{i\ep}\Sigma(i\ep)\Big)\\
&= \fint_{-\infty}^\infty \fract{d\ep}{2\pi}\,\int_{-\infty}^\infty \fract{d\ep'}{2\pi}\,
\Tr\Big(\delta G(i\ep)\cdot \Gamma_0^S(i\ep,i\ep')\tdot \partial_{i\ep'}G(i\ep')\Big)\,.
\label{second step}
\eal
It is important to notice that the derivation of Eq.~\eqn{simple WT identity} through the 
Ward-Takahashi identity implies that the integral over $\ep'$ in \eqn{second step} is not the 
Cauchy principal value at $\ep'=0$. In other words, $\partial_{i\ep'}G(i\ep')$ contains 
the $\delta$-function term $2 G_2(0^+)\,\delta(\ep')$ that exists in metals. \\
Considering $\delta I_L^{(2)}$, upon integration by parts, taking into account the possible discontinuity at $\ep=0$, and recalling that the functional derivative $\delta\Sigma/\delta G$ 
is the irreducible vertex $\Gamma_0$, which coincides with $\Gamma^S_0$ at zero transferred frequency and momentum, we find 
\beal
\delta I_L^{(2)} &= - \fint_{-\infty}^\infty \fract{d\ep}{2\pi}\,
\Tr\Big(\partial_{i\ep} G(i\ep)\cdot \delta\Sigma(i\ep)\Big)\\
&\qquad 
-\fract{1}{\pi}\,\bigg\{\Tr\Big(G_2(0^+)\,\delta\Sigma_1(0^+)\Big)
+ \Tr\Big(G_1(0^+)\,\delta\Sigma_2(0^+)\Big)\bigg\}\\
&= - \int_{-\infty}^\infty \fract{d\ep}{2\pi}\,
\Tr\Big(\partial_{i\ep} G(i\ep)\cdot \delta\Sigma(i\ep)\Big)
-\fract{1}{\pi}\,\Tr\Big(G_1(0^+)\,\delta\Sigma_2(0^+)\Big)\\
&= - \int_{-\infty}^\infty \fract{d\ep}{2\pi}\,\fint_{-\infty}^\infty\fract{d\ep'}{2\pi}
\Tr\Big(\partial_{i\ep} G(i\ep)\cdot \Gamma_0^S(i\ep,i\ep')\tdot \delta G(i\ep')\Big)\\
&\qquad -\fract{1}{\pi}\,\Tr\Big(G_1(0^+)\,\delta\Sigma_2(0^+)\Big)\,.
\label{third step}
\eal
Summing \eqn{second step} and \eqn{third step} we finally obtain 
\be
\delta I_L = -\fract{1}{\pi}\,\Tr\Big(G_1(0)\,\delta\Sigma_2(0^+)\Big)\,.
\label{IL is a topological invariant}
\ee
It follows that $\delta I_L = 0$ whenever $\delta\Sigma_2(0^+)\equiv 0$. Hence, metals with $\Sigma_2(0^+)=0$ and insulators without a Luttinger surface \cite{Igor-PRB2003} keep the same $I_L$ under any variation $G(i\ep)\to G(i\ep)+\delta G(i\ep)$ such that the self-energy is continuous at $\ep=0$ both before and after the variation. \\
By contrast, $I_L$ can change only if the self-energy variation jumps at $\ep=0$. At $T=0$ and for a non-degenerate ground state, this generically requires a singular $\delta\Sigma_2(0^+)$, i.e. a pole of $\Sigma(i\ep)$ at $\ep=0$ and thus a vanishing $\det G(0)$. In periodic systems, this corresponds to the appearance of a Luttinger surface, the location in the Brillouin zone of the zeros of $\det G(0,\bk)$; in finite systems, to isolated points in parameter space where $I_L$ jumps by an integer.\\
Once a Luttinger surface is present, in either a metal or an insulator, any variation that changes its shape in momentum space yields, through \eqn{IL is a topological invariant}, a $\delta I_L$ that is in general non-zero, so that $I_L$ varies. \\

\noindent
The above invariance conditions of $I_L$ under a variation $\delta G(i\ep)$ can be 
given an enlightening geometrical representation.
Through \eqn{LT invariance T=0}, the Luttinger integral can be written as:
\beal
I_L &= \fint_{-\infty}^\infty \fract{d\ep}{2\pi}\,
\Tr\Big(G(i\ep)\cdot\partial_{i\ep}\Sigma(i\ep)\Big) \\
&= -\fract{1}{\pi}\,\Tr\Big(\Sigma_1(0)\,G_2(0^+)\Big)-\fint_{-\infty}^\infty \fract{d\ep}{2\pi}\,
\Tr\Big(\Sigma(i\ep)\cdot\partial_{i\ep}G(i\ep)\Big)\\
&= -\fract{1}{\pi}\,\Tr\Big(\Sigma_1(0)\,G_2(0^+)\Big) -\fint_{-\infty}^\infty \fract{d\ep}{2\pi}\,\bigg\{
\Tr\Big(\Sigma_1(\ep)\cdot\partial_{\ep}G_2(\ep)\Big)
+ \Tr\Big(\Sigma_2(\ep)\cdot\partial_{\ep}G_1(\ep)\Big)\bigg\}\,.
\label{conclusions 1}
\eal
If we consider physical variations of the Green's function $\delta G(i\ep) = \delta G_1(\ep) + i \delta G_2(\ep)$, where $\delta G_1(\epsilon)$ and $\delta G_2(\epsilon)$ inherit the symmetry properties of $G_1(\epsilon)$ and $G_2(\epsilon)$ (i.e., even and odd under $\epsilon \rightarrow -\epsilon$, respectively), we find:
\bealn
\delta \Phi[G] &= \int_{-\infty}^\infty \fract{d\ep}{2\pi}\,
\Tr\Big(\Sigma(i\ep)\cdot\delta G(i\ep)\Big)
= \int_{-\infty}^\infty \fract{d\ep}{2\pi}\,\bigg\{
\Tr\Big(\Sigma_1(\ep)\cdot\delta G_1(\ep)\Big)
- \Tr\Big(\Sigma_2(\ep)\cdot\delta G_2(\ep)\Big)\bigg\}\\
&\qquad + i\,\int_{-\infty}^\infty \fract{d\ep}{2\pi}\,\bigg\{
\Tr\Big(\Sigma_1(\ep)\cdot\delta G_2(\ep)\Big)
+ \Tr\Big(\Sigma_2(\ep)\cdot\delta G_1(\ep)\Big)\bigg\}\\
&= \int_{-\infty}^\infty \fract{d\ep}{2\pi}\,\bigg\{
\Tr\Big(\Sigma_1(\ep)\cdot\delta G_1(\ep)\Big)
- \Tr\Big(\Sigma_2(\ep)\cdot\delta G_2(\ep)\Big)\bigg\}\,.
\eal
This structure implies the exact functional derivative relations:
\beal
\fract{\delta \Phi[G]}{\delta G_1(\ep)} &\equiv \Sigma_1(\ep)\,,&
\fract{\delta \Phi[G]}{\delta G_2(\ep)} &\equiv -\Sigma_2(\ep)\,.
\label{conclusions 2}
\eal
\begin{figure}[ht]
\centerline{\includegraphics{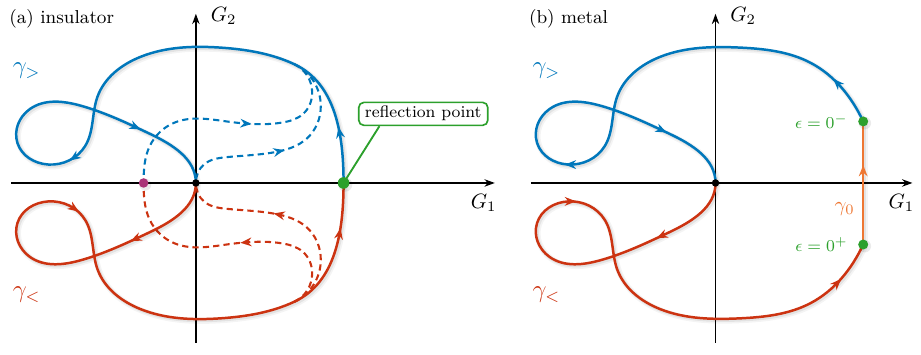}}
\caption{Sketch of a generic path traversed by $G=G_1+i\,G_2$ along the Matsubara trajectory from $\ep=+\infty$ to $\ep=-\infty$ in the case of a correlated insulator, panel (a), 
and a correlated metal, panel (b).  
Even though $G_1$ and $G_2$ are matrices that reside in a multidimensional space, for simplicity we project the path onto two dimensions as if they were scalars. 
The path is composed of three segments: $\gamma_<$, in red, which is restricted to the lower half-plane ($G_2<0$) and corresponds to the trajectory from $\ep=+\infty$ to $\ep=0^+$, 
its reflected path $\gamma_>$ ($G_2\to-G_2$), in blue, and the vertical segment $\gamma_0$, in orange, 
that connects the endpoint $(G_1(0),G_2(0^+))$ of $\gamma_<$ to the starting point $(G_1(0),G_2(0^-))=(G_1(0),-G_2(0^+))$ of $\gamma_>$. In the insulator, panel (a), $\gamma_0$ shrinks to a point. The crucial element along the path is the reflection point (indicated in green), which corresponds to $(G_1(0),0)$. Specifically, $I_L$ can change only if the reflection point reaches and subsequently crosses the origin $(G_1,G_2)=(0,0)$ (dashed segments in panel (a)) carrying the topological defect with it.}
\label{path}
\end{figure}
By substituting \eqn{conclusions 2} into \eqn{conclusions 1}, the Luttinger integral 
can be compactly recast as:
\beal
I_L &= -\fint_{-\infty}^\infty \fract{d\ep}{2\pi} \left( \Tr\bigg( \fract{\delta \Phi[G]}{\delta G_1(\ep)} \cdot \partial_\ep G_2(\ep) \bigg) 
- \Tr\bigg( \fract{\delta \Phi[G]}{\delta G_2(\ep)} \cdot \partial_\ep G_1(\ep) \bigg) \right)  \\
&\qquad -\fract{1}{\pi}\,\Tr\Big(\Sigma_1(0)\,G_2(0^+)\Big)\\
&= \fint_{+\infty}^{-\infty} \fract{d\ep}{2\pi} \left( \Tr\bigg( \fract{\delta \Phi[G]}{\delta G_1(\ep)} \cdot \partial_\ep G_2(\ep) \bigg) 
- \Tr\bigg( \fract{\delta \Phi[G]}{\delta G_2(\ep)} \cdot \partial_\ep G_1(\ep) \bigg) \right)  \\
&\qquad +\fract{1}{2\pi}\,\Tr\bigg(\Sigma_1(0)\,\Big(G_2(0^-)-G_2(0^+)\Big)\bigg)\\
&= \fract{1}{2\pi}\,\oint_\gamma \Bigg\{ \Tr\bigg( \fract{\delta \Phi[G]}{\delta G_1} \cdot dG_2 \bigg) 
- \Tr\bigg( \fract{\delta \Phi[G]}{\delta G_2} \cdot dG_1 \bigg) \Bigg\} \,,
\label{monodromy integral}
\eal
where $\gamma=\gamma_<+\gamma_0+\gamma_>$. Here $\gamma_<$ is the Matsubara trajectory from $\ep=+\infty$ to $\ep=0^+$, $\gamma_>$ the one from $\ep=0^-$ to $\ep=-\infty$, and $\gamma_0$ the vertical segment that joins $\big(G_1(0),G_2(0^+)\big)$ to $\big(G_1(0),G_2(0^-)\big)=\big(G_1(0),-G_2(0^+)\big)$. Indeed, along $\gamma_0$, $\delta\Phi[G]/\delta G_1=\Sigma_1(0)$ and $dG_1=0$, so that the line integral of the one-form along $\gamma_0$ is precisely the boundary term in the second equation of \eqn{monodromy integral}. Equation~\eqn{monodromy integral} thus demonstrates that $I_L$ is the line integral of a functional one-form derived from $\Phi[G_1,G_2]$, specifically the conjugate differential form $d^c\Phi[G_1,G_2]$, over the closed path $\gamma$.\\
In an insulator, $G_2(0^\pm)=0$ and the segment $\gamma_0$ shrinks to a point. The trajectory $\gamma_<$ begins at $(G_1,G_2)=(0,0^-)$, which corresponds to $\ep=\infty$, traverses entirely the lower half-plane ($G_2<0$), and ends at the reflection point $(G_1,G_2)=(G_1(0),0)$ when $\ep=0^+$. The second half of the path, $\gamma_>$, is obtained by reflecting the first half, $G_1\to G_1$ and $G_2\to-G_2$. It starts at $\ep=0^-$, where once again $(G_1,G_2)=(G_1(0),0)$, and terminates at $(G_1,G_2)=(0,0^+)$, implying $\ep=-\infty$. Because physical boundary conditions require $G(i\infty)=G(-i\infty)=0$, the Matsubara trajectory from $\ep=\infty$ to $\ep=-\infty$ natively forms a closed loop in the space of Green's functions, see Fig.~\ref{path}(a). \\
In a metal, $\ep=0^+$ corresponds to $\big(G_1(0),G_2(0^+)\big)$ and $\ep=0^-$ to $\big(G_1(0),-G_2(0^+)\big)$, so that $\gamma_<$ and $\gamma_>$ apparently do not meet. However, they are joined by the segment $\gamma_0$, whose contribution is the boundary term in \eqn{monodromy integral}, which is now finite. The path is therefore closed also in a metal, see Fig.~\ref{path}(b). We note that $\gamma_0$ crosses the real axis at $(G_1(0),0)$, i.e. at the reflection point of the insulating case. Moreover, since $\gamma_>$ is the mirror image of $\gamma_<$, then 
\bealn
\fract{\delta \Phi[G]}{\delta G_1} &\xrightarrow[\gamma_<\,\to\,\gamma_>]{} \fract{\delta \Phi[G]}{\delta G_1}\;,&
\fract{\delta \Phi[G]}{\delta G_2} &\xrightarrow[\gamma_<\,\to\,\gamma_>]{} -\fract{\delta \Phi[G]}{\delta G_2}\;,\\
dG_{1} &\xrightarrow[\gamma_<\,\to\,\gamma_>]{} -dG_{1}\,,&
dG_{2} &\xrightarrow[\gamma_<\,\to\,\gamma_>]{} dG_{2}\,,
\eal
which implies that
\be
\fract{1}{2\pi}\int_{\gamma_>}\Tr\Big(d^c\Phi[G_1,G_2]\Big)
= \fract{1}{2\pi}\int_{\gamma_<}\Tr\Big(d^c\Phi[G_1,G_2]\Big)\,,
\label{mirror}
\ee
namely, that $\gamma_>$ and $\gamma_<$ add together constructively, in contrast to $\Tr\big(d\Phi[G_1,G_2]\big)$, for which they cancel each other. The segment $\gamma_0$ is mapped onto itself, and its two halves, above and below the $G_1$-axis, also contribute equally.\\

\noindent
From Eq.~\eqn{IL is a topological invariant} we thus conclude that $I_L$, which is an integer since both $N$ and $N_L$ at $T=0$ are, remains invariant under any continuous deformation of the path $\gamma$ along which the self-energy stays continuous at $\ep=0$. At $T=0$ this amounts to requiring that $G_1(0)$ and $G_2(0^+)$ never share a null vector, i.e. that no reflection point passes through the origin, see Fig.~\ref{path}(a),  whereas passing through infinity, namely changing the Fermi surface, is allowed.\\ 
$I_L$ is therefore a topological invariant in the following sense: it is an integer, and it remains constant along any continuous path in the Hamiltonian parameter space along which $\det G(i0^+)$ does not vanish. When $\det G(i0^+)$ happens to vanish, $I_L$ loses its protection and may vary while $\det G(i0^+)$ stays zero. Once $\det G(i0^+)$ becomes again nonzero and $\Sigma(i\ep)$ regains analyticity everywhere, including at $\ep=0$, $I_L$ has changed to a different quantized integer value. This observation is consistent with the example discussed at the end of Sec.~\ref{Final remarks}, and points out that only when $I_L$ changes can one find signals in the local-in-$\ep$ behavior of $\Sigma(i\ep)$. Once $I_L$ has changed, the anomaly becomes a global property of the LW functional unrelated to any infrared singularity.

\FloatBarrier
\section{A few relevant examples}
\label{A few relevant examples}

Here we briefly examine a few simple examples of finite systems that implement the ideas from the prior Sec.~\ref{The Luttinger Integral as a novel topological invariant}.

\subsection{Two-orbital toy model}
\label{Two-orbital toy model}

When the interaction is turned on, at first it modifies the non-interacting ground state by transferring pairs of electrons from occupied to unoccupied single-particle levels. To mimic  this effect, we consider a very simple model that consists of just two orbitals, 1 and 2, 
described by the Hamiltonian
\be
H = \sum_{\sigma=\up,\down}\, \Big(\ep_1\,c^\dagger_{1\sigma}\,c^\dagga_{1\sigma}
+ \ep_2\,c^\dagger_{2\sigma}\,c^\dagga_{2\sigma}\Big)-J\, \Big(c^\dagger_{1\up}\,c^\dagger_{1\down}\;c^\dagga_{2\down}\,c^\dagga_{2\up}
+c^\dagger_{2\up}\,c^\dagger_{2\down}\;c^\dagga_{1\down}\,c^\dagga_{1\up}\Big)\,.
\label{toy Ham}
\ee
Without loss of generality, we assume that $\ep_1<\ep_2$. The lowest energy state with two electrons 
has energy 
\be
E_0 = \ep_1+\ep_2 -\sqrt{\big(\ep_2-\ep_1\big)^2 + J^2\;} \coloneq \ep_1+\ep_2 - \Delta\,,
\label{toy E0}
\ee
and reads
\be
\ket{\psi_0} = \cos\theta\, c^\dagger_{1\up}\,c^\dagger_{1\down}\ket{0}
+ \sin\theta\, c^\dagger_{2\up}\,c^\dagger_{2\down}\ket{0}\,,
\label{toy psi0}
\ee
where here and in the next section $\ket{\!0}$ denotes the vacuum state, and 
\bealn
\cos{2\theta} &= \fract{\;\ep_2-\ep_1\;}{\Delta}\;,& 
\sin{2\theta} &= \fract{\;J\;}{\Delta}\;.
\eal
We further assume that $\ket{\psi_0}$ is the ground state, which implies the following inequality
\be
\Delta > \max\big(-\ep_1,\ep_2,-\ep_1-\ep_2,\ep_1+\ep_2\big)\,.
\label{toy constraints}
\ee
The Green's function is diagonal in the orbital index, is spin independent, and its matrix elements read: 
\beal
G_{11}(i\ep) &= 
\fract{\;i\ep -2\ep_2+\ep_1}{\;\big(i\ep-\ep_2+\Delta\big)
\big(i\ep-\ep_2-\Delta\big)\;}\;,&
G_{22}(i\ep) &= \fract{\;i\ep-2\ep_1+\ep_2\;}
{\;\big(i\ep-\ep_1+\Delta\big)\big(i\ep-\ep_1-\Delta\big)\;}\;, 
\label{toy G}
\eal
and correspond to a diagonal self-energy with spin-independent elements   
\beal
\Sigma_{11}(i\ep) &= \fract{J^2}{\;i\ep -2\ep_2+\ep_1\;}\;,&
\Sigma_{22}(i\ep) &= \fract{J^2}{\;i\ep-2\ep_1+\ep_2\;}\;.
\label{toy self-energies}
\eal
\begin{figure}[ht]
\centerline{\includegraphics{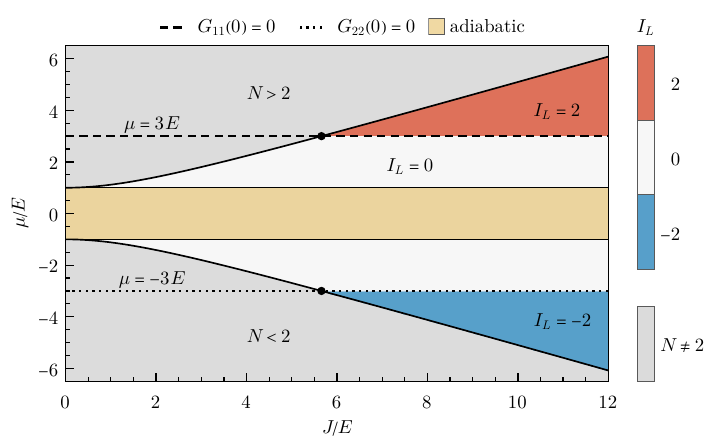}}
\caption{Luttinger integral corresponding to the Hamiltonian \eqn{toy Ham} with $\ep_1=-E-\mu$ and $\ep_2=E-\mu$ ($E>0$)
in the $\mu/E$ vs. $J/E$ parameter space where the ground state has two electrons, $N=2$ ($N\not=2$ in the gray regions). The horizontal lines $\mu=3E$ and $\mu=-3E$ correspond to the zeros of $G_{11}(0)$ and $G_{22}(0)$, respectively.}
\label{toy-pd}
\end{figure}
It is easy to show that 
\beal
I_L &= \sum_{\sigma=\up,\down}\,\sum_{i=1}^2\, \fint_{-\infty}^\infty \fract{d\ep}{2\pi}\;G_{ii}(i\ep)\;\fract{\partial
\Sigma_{ii}(i\ep)}{\partial i\ep} \equiv \fint_{-\infty}^\infty \fract{d\ep}{2\pi}\,
\fract{\partial \,\Tr\big(\ln X(i\ep)\big)}{\partial i\ep}\,,
\label{1st example IL} 
\eal
where one readily finds that the matrix $X(i\ep)$ is also diagonal with 
elements
\beal
X_{1\sigma}(i\ep) &= \big(i\ep -2\ep_2+\ep_1\big)\;
\big(i\ep-\ep_2+\Delta\big)^{(\ep_2-\ep_1-\Delta)/2\Delta}\;
\big(i\ep-\ep_2-\Delta\big)^{(\ep_1-\ep_2-\Delta)/2\Delta}\;,\\
X_{2\sigma}(i\ep) &= \big(i\ep -2\ep_1+\ep_2\big)\;
\big(i\ep-\ep_1+\Delta\big)^{(\ep_1-\ep_2-\Delta)/2\Delta}\;
\big(i\ep-\ep_1-\Delta\big)^{(\ep_2-\ep_1-\Delta)/2\Delta}\;.
\eal
The constraint \eqn{toy constraints} implies 
\bealn
\Delta-\ep_1 &> \Delta-\ep_2 >0\,,& \Delta+\ep_2 &>\Delta+\ep_1 >0\,, 
\eal
so that \eqn{1st example IL} becomes simply
\be
I_L = 2\Big(1 -\theta\big(2\ep_2-\ep_1\big) -\theta\big(2\ep_1-\ep_2\big)\Big)\,.
\label{1st example IL explicit}
\ee
To simplify the graphical representation of $I_L$, we write $\ep_1=-E-\mu$, $\ep_2=E-\mu$ 
with $0<E\equiv 1$ the unit of energy. It follows that $2\ep_2-\ep_1\equiv 3-\mu$, 
$2\ep_1-\ep_2\equiv -3-\mu$, thus
\be
I_L = 2\Big(1 -\theta\big(3-\mu\big) -\theta\big(-3-\mu\big)\Big)\,,
\label{1st example IL explicit bis}
\ee
with the constraint 
\be
\Delta = \sqrt{4+J^2\;} > \max\big(1+\mu,1-\mu,2\mu,-2\mu\big)\,.
\label{toy constraints bis}
\ee
In Fig.~\ref{toy-pd} we plot $I_L$ in the region where the ground state has two electrons, 
i.e. when the inequality \eqn{toy constraints bis} is valid. We observe that Luttinger's theorem may hold even when the non-interacting and interacting ground states are not adiabatically connected.

\FloatBarrier
\subsection{Hubbard dimer}
\label{Hubbard dimer}

We consider now a Hubbard dimer with Hamiltonian
\be
H = -t\,\sum_\sigma\, \big(c^\dagger_{1\sigma}\,c^\dagga_{2\sigma} + c^\dagger_{2\sigma}\,c^\dagga_{1\sigma}\big) -\mu\,\sum_{i=1}^2\,\sum_\sigma\, n_{i\sigma} 
+ U\,\sum_{i=1}^2\, n_{i\up}\,n_{i\down}\,,
\label{Ham dimer}
\ee
with both $t$ and $U$ positive.
The Hamiltonian \eqn{Ham dimer} is invariant under $1\leftrightarrow 2$, which implies that the 
bonding $b$ and antibonding $a$ combinations
\beal
c^\dagga_{b\sigma} &=\fract{1}{\sqrt{2}}\,\big(c^\dagga_{1\sigma} + c^\dagga_{2\sigma}\big)\,,&
c^\dagga_{a\sigma} &=\fract{1}{\sqrt{2}}\,\big(c^\dagga_{1\sigma} - c^\dagga_{2\sigma}\big)\,,
\label{dimer bonding antibonding}
\eal
are irreducible representations of the symmetry group. \\
The lowest energy state with two electrons is 
\be
\ket{\psi_0} = \cos\theta\, c^\dagger_{b\up}\,c^\dagger_{b\down}\ket{0}  -
\sin\theta\, c^\dagger_{a\up}\,c^\dagger_{a\down}\ket{0}\,,
\label{dimer psi0}
\ee
with energy 
\be
E_0 = -2\mu +\fract{U}{2} -\sqrt{\left(\fract{U}{2}\right)^2 +4t^2\;} = -2\mu +\fract{U}{2} -\Delta\,,
\label{dimer E0}
\ee
and where 
\bealn
\cos 2\theta &= \fract{\;2t\;}{\Delta}\;,& \sin 2\theta &= \fract{U}{\;2\Delta\;}\;.
\eal
The state $\ket{\psi_0}$ in \eqn{dimer psi0} is the actual ground state if 
\be
\mu_\text{min}(t,U) \leq \mu \leq \mu_\text{max}(t,U)\,,
\label{dimer condition mu}
\ee
where 
\bealn
\mu_\text{min}(t,U) &= \fract{U}{2}+t -\Delta\,,&
\mu_\text{max}(t,U) &= \fract{U}{2}-t +\Delta\,.
\eal
In particular, the non-interacting and interacting ground states are adiabatically connected if 
$\mu_\text{min}(t,U) \leq \mu \leq \mu_{\text{max}}(t,0)= t$. 
\begin{figure}[ht]
\centerline{\includegraphics{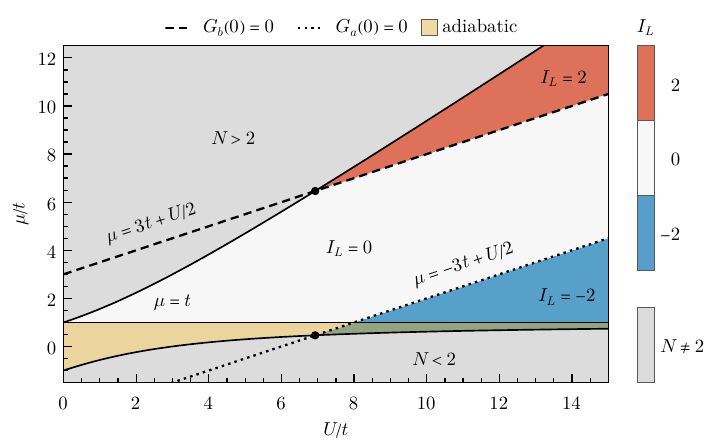}}
\caption{Luttinger integral corresponding to the Hamiltonian \eqn{Ham dimer} in the $\mu$ vs. $U$ parameter space where the ground state has two electrons. The dashed and dotted lines correspond to the zeros of $G_b(0)$ and $G_a(0)$, respectively.}
\label{dimer-pd}
\end{figure}
Assuming \eqn{dimer condition mu}, the Green's function, which is diagonal in the bonding-antibonding 
basis \eqn{dimer bonding antibonding}, has elements
\beal
G_b(i\ep) &= \fract{\sin^2\theta}{\;i\ep +\mu -2t -\mu_\text{max}(t,U)\;}
+\fract{\cos^2\theta}{\;i\ep +\mu -\mu_\text{min}(t,U)\;}\\
&= \fract{i\ep+\mu-3t-U/2}{\;\big(i\ep +\mu -2t -\mu_\text{max}(t,U)\big)
\big(i\ep +\mu -\mu_\text{min}(t,U)\big)\;}
\;,\\
G_a(i\ep) &= \fract{\cos^2\theta}{\;i\ep +\mu -\mu_\text{max}(t,U)\;}
+\fract{\sin^2\theta}{\;i\ep +\mu +2t -\mu_\text{min}(t,U)\;}\\
&= \fract{i\ep+\mu+3t-U/2}{\;\big(i\ep +\mu -\mu_\text{max}(t,U)\big)
\big(i\ep +\mu +2t -\mu_\text{min}(t,U)\big)\;}\;.
\eal
We observe that $G_b(0)>0$ if $\mu<3t+U/2$, otherwise $G_b(0)\leq0$. Similarly, 
$G_a(0)<0$ if $\mu>-3t+U/2$, otherwise $G_a(0)\geq 0$. 
In Fig.~\ref{dimer-pd} we show the value of the Luttinger integral corresponding to the Hamiltonian \eqn{Ham dimer} in the parameter space where the ground state has two electrons, i.e. when Eq.~\eqn{dimer condition mu} holds. We observe that if 
$\mu_\text{min}(t,U) \leq \mu \leq t$, i.e. when the non-interacting and interacting ground states are adiabatically connected, Luttinger's theorem can be still violated, the blue region with $I_L=-2$ 
where $\mu_\text{min}(t,U) \leq \mu \leq -3t +U/2$.

\FloatBarrier
\subsection{Hubbard plaquette pierced by a magnetic flux}
\label{Hubbard plaquette pierced by a magnetic flux}

\begin{figure}[ht]
\centerline{\includegraphics{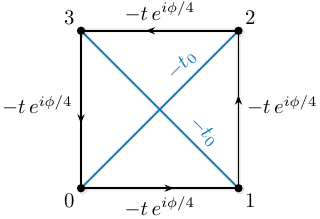}}
\caption{Tight-binding model with nearest, $-t$, and next-nearest, $-t_0$, neighbor hopping on a plaquette pierced by a magnetic flux $\phi$.}
\label{plaquette}
\end{figure}
The next example we consider is that of a single-orbital Hubbard model on a plaquette pierced by a magnetic flux $\phi$ and with nearest and next-nearest neighbor hopping, see Fig.~\ref{plaquette}. This is a very simple model where we can investigate the validity of Luttinger's theorem with broken time-reversal symmetry. The Hamiltonian reads
\be
H = H_0 -\mu\,\sum_{i=0}^3\,\sum_{\sigma=\up,\down}\,n_{i\sigma} 
+ U\,\sum_{i=0}^3\,n_{i\up}\,n_{i\down}\,,
\label{plaquette Ham}
\ee
where $n_{i\sigma} = c^\dagger_{i\sigma}\,c^\dagga_{i\sigma}$ is the occupation number of spin 
$\sigma=\up,\down$ electrons on site $i=0,\dots,3$, and the non-interacting Hamiltonian is, see Fig.~\ref{plaquette},     
\beal
H_0 &= -t\,\sum_\sigma\,\sum_{i=0}^3\,
\Big(\esp{i\phi/4}\;c^\dagger_{i\sigma}\,c^\dagga_{i+1\sigma} + \esp{-i\phi/4}\;c^\dagger_{i+1\sigma}\,c^\dagga_{i\sigma}\Big)\\
&\qquad\qquad\quad 
- t_0\,\sum_\sigma\, \Big(c^\dagger_{2\sigma}\,c^\dagga_{0\sigma} + 
c^\dagger_{0\sigma}\,c^\dagga_{2\sigma} +
c^\dagger_{1\sigma}\,c^\dagga_{3\sigma} + 
c^\dagger_{3\sigma}\,c^\dagga_{1\sigma}\Big)\,,
\label{plaquette Ham 0}
\eal
where $c^\dagga_{4\sigma}\equiv c^\dagga_{0\sigma}$.
\begin{figure}[ht]
\centerline{\includegraphics{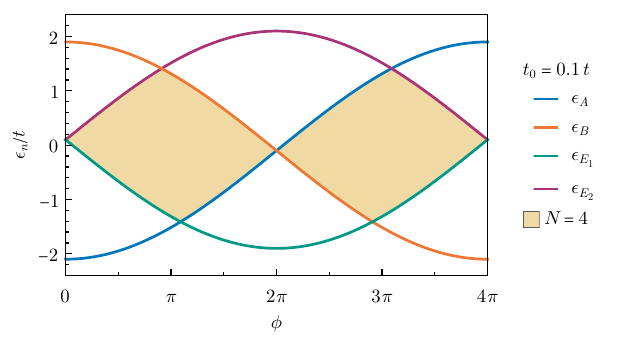}}
\caption{Non-interacting eigenvalues, $\ep_n(\phi)$, $n=A,B,E_1, E_2$, as function of $\phi$. The values of $\mu$ such that the non-interacting ground state has $N=4$ electrons at flux $\phi$ are marked by a yellow region.}
\label{H0-01}
\end{figure}
\begin{figure}[bht]
\centerline{\includegraphics{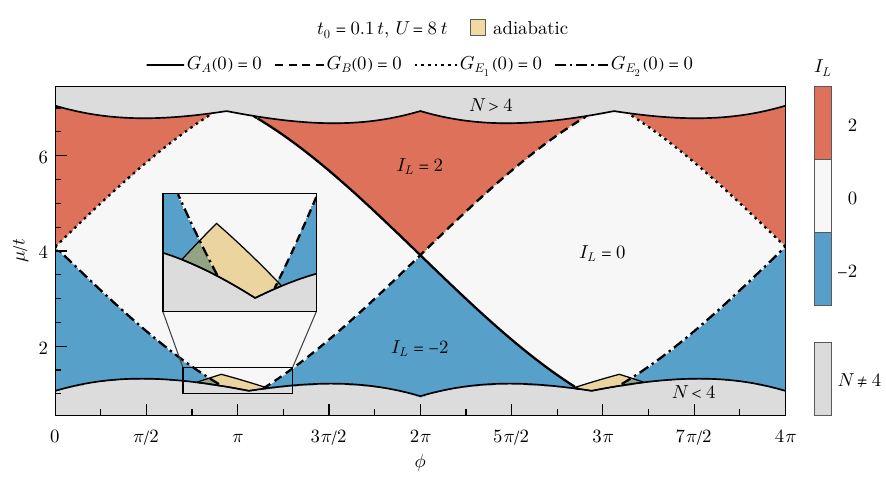}}
\caption{Value of the Luttinger integral $I_L$ of the plaquette pierced by a magnetic flux as function of $\phi$ and $\mu$ at fixed $t_0=0.1t$ and $U=8t$ where the ground state has four electrons. The black lines correspond to the zeros of the Green's function at frequency zero. The region where the ground state is adiabatically connected to the non-interacting ground state is colored in yellow.}
\label{IL_plaquette_U8}
\end{figure}
At finite flux, the Hamiltonian \eqn{plaquette Ham} is only invariant under the point group $C_4$ that has four one-dimensional irreducible representations (irreps); two real ones, $A$ and $B$, respectively, even and odd under $C_4$, and two complex ones, $E_1$ and $E_2$. \\
The eigenvalues of $H_0$ can be readily obtained:
\beal
\ep_A(\phi) &= -t_0 - 2t\,\cos\fract{\phi}{4}
\;,&
\ep_B(\phi) &= -t_0 +2t\,\cos\fract{\phi}{4}\;,\\
\ep_{E_1}(\phi) &= t_0 - 2t\,\sin\fract{\phi}{4}
\;,&
\ep_{E_2}(\phi) &= t_0 +2t\,\sin\fract{\phi}{4}\;,
\eal
where the states are labeled as the corresponding irreps, 
and are periodic in $\phi$ with period $4\pi$.
Their evolution with $\phi$ is non-trivial because of several level crossings, see Fig.~\ref{H0-01}. Hereafter, we set $t=1$ as the unit of energy, and assume $0\leq t_0\leq t=1$.\\

\begin{figure}[bht]
\centerline{\includegraphics{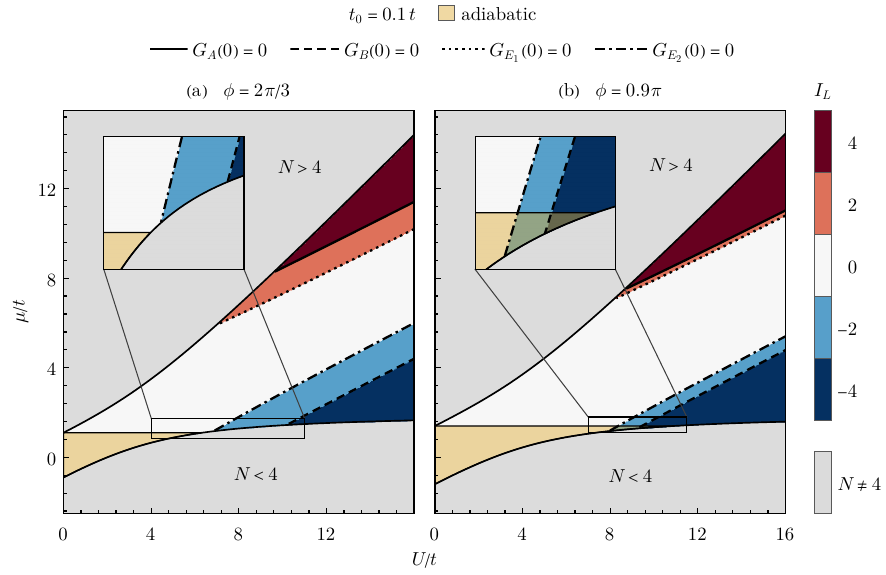}}
\caption{Value of the Luttinger integral $I_L$ of the plaquette pierced by a magnetic flux as function of $U$ and $\mu$ at fixed $t_0=0.1t$ and $\phi=2\pi/3$ [panel (a)] and $\phi=0.9\pi$ [panel (b)]  where the ground state has four electrons. The black lines correspond to the zeros of the Green's function at frequency zero. The yellow region corresponds to the case in which non-interacting and interacting ground states are adiabatically connected.}
\label{IL_plaquette_Uvsmu}
\end{figure}

\noindent
The lowest energy eigenstate with $N=4$ electrons is non-degenerate and resides in the spin-singlet subspace of $B$ symmetry for any $\phi$ at $U>0$, and only for 
$\phi\not=0,2\pi,4\pi$ at $U=0$.\\
In Fig.~\ref{IL_plaquette_U8} we show the Luttinger integral as function of $\phi$ and $\mu$ at fixed $t_0=0.1t$ and $U=8t$. The region where non-interacting and interacting ground states are adiabatically connected is marked in yellow, see also the inset.
It is interesting to note that the shape of the region where $I_L=0$ is similar to the region where the non-interacting ground state has four electrons (yellow region in Fig.~\ref{H0-01}).
To further illustrate the role of adiabatic continuity, Fig.~\ref{IL_plaquette_Uvsmu} shows the
$U$ vs.~$\mu$ phase diagram at two fixed values of the flux and $t_0=0.1t$. The shaded yellow area marks the region in which
the interacting ground state is adiabatically connected to the non-interacting. For $\phi=2\pi/3$ [panel (a)] this region lies entirely within $I_L=0$, so Luttinger's theorem holds
wherever the ground state is adiabatically connected to the non-interacting one. For $\phi=0.9\pi$
[panel (b)], instead, the shaded region extends into the area with $I_L\neq0$.

\FloatBarrier
\section{Conclusions}
\label{Conclusions}

In this work, we have provided a non-perturbative resolution of the long-standing apparent paradox surrounding Luttinger's theorem violations in strongly correlated systems. By re-examining the properties of the Luttinger-Ward functional $\Phi[G]$ under imaginary frequency translations, we demonstrated that reported violations of Luttinger's theorem imply neither that $\Phi[G]$ loses its invariance under frequency shifts nor that the skeleton diagram representation of physical quantities breaks down. In fact, we have shown that this representation, regardless of potential failures of the skeleton diagrammatic \emph{expansion}, remains consistent with the thermodynamic Streda formula. \\

\noindent
Specifically, we find that the Green's function $G(i\ep,\bk)=G(-i\ep,\bk)^\dagger=G_1(\ep,\bk)+i\,G_2(\ep,\bk)$ along the Matsubara trajectory from 
$\ep=+\infty$ to $\ep=-\infty$ traverses a closed path $\gamma$ in the $(G_1,G_2)$ space, 
as the ones sketched in Fig.~\ref{path} for correlated insulators and metals, 
panel (a) and (b), respectively. One can readily demonstrate that, if $L$ is the total number of available single-particle states ($L=MV$ in a periodic model with $V$ unit cells and $M$ Bloch waves, spin included) then the Luttinger count \eqn{Luttinger's theorem} can be rewritten as\footnote{
Since 
\bealn
\sum_{\ell=1}^L\, \langle c^\dagga_\ell\,c^\dagger_\ell\rangle
= \sum_{\ell=1}^L\, \big(1-\langle c^\dagger_\ell\,c^\dagga_\ell\rangle\big) = L -N\,,
\eal
similarly, if 
\bealn
N_L &= -\fint_{-\infty}^\infty \fract{d\ep}{2\pi}\;\esp{i\ep 0^+}\;
\fract{\partial}{\partial i\ep}\, \Tr\Big(\ln G(i\ep)\Big) = 
-\fint_{-\infty}^\infty \fract{d\ep}{2\pi}\;\esp{i\ep 0^+}\;
\fract{\partial}{\partial i\ep}\, \Tr\Big(\ln G_*(i\ep)\Big)\,,
\eal
then
\bealn
N_L -L &= -\fint_{-\infty}^\infty \fract{d\ep}{2\pi}\;\esp{-i\ep 0^+}\;
\fract{\partial}{\partial i\ep}\, \Tr\Big(\ln G(i\ep)\Big)\,,
\eal
which implies that 
\bealn
N_L &= -\fint_{-\infty}^\infty \fract{d\ep}{2\pi}\;\cos(\ep\, 0^+)\;
\fract{\partial}{\partial i\ep}\, \Tr\Big(\ln G(i\ep)\Big)
+\fract{L}{2}= \fint_{\infty}^{-\infty} \fract{d\ep}{2\pi i}\;
\fract{\partial}{\partial \ep}\, \Tr\Big(\ln G(i\ep)\Big)
+\fract{L}{2}\;,
\eal
where we safely drop the cosine term, thus Eq.~\eqn{Luttinger count}. 
} 
\be
N_L = \fract{1}{2\pi i}\,\oint_\gamma d\, \Tr\big(\ln G\big) + \fract{L}{2}\,,
\label{Luttinger count}
\ee
where the trace includes summation over momentum, Bloch waves and spin.
We have shown that the Luttinger integral $I_L$ is the circulation of the 
conjugate differential form $d^c\Phi[G_1,G_2]$ of the Luttinger-Ward functional along the same closed path, namely, via Eq.~\eqn{monodromy integral}:
\be
I_L = \fract{1}{2\pi}\,\oint_\gamma \Bigg\{ \Tr\bigg( \fract{\delta \Phi[G]}{\delta G_1} \cdot dG_2 \bigg) 
- \Tr\bigg( \fract{\delta \Phi[G]}{\delta G_2} \cdot dG_1 \bigg) \Bigg\}\,.
\label{Luttinger integral}
\ee
Also $I_L$ is quantized in integer values and remains strictly invariant under any smooth deformation of path $\gamma$. Consequently, $I_L$ vanishes, and Luttinger's theorem holds, within a finite parameter region adjacent to the non-interacting case, where $I_L$ is simply zero. We remark that, see the examples discussed in Sec.~\ref{A few relevant examples}, the validity of the theorem does \emph{not} reduce strictly to the fact that the interacting ground state can be adiabatically continued into the non-interacting one.\\ 
The Luttinger integral $I_L$ can become non-zero, and Luttinger's theorem violated, only if 
$\text{det}\big(G(i0^+)\big)$ vanishes. This signals a divergence of the self-energy at $\ep=0$. Once the topological charge has changed, the self-energy can become analytic again, including at $\ep=0$. Because of this divergence of the self-energy, the violation of the Luttinger theorem has often been associated with a breakdown of the skeleton \emph{expansion}. Indeed, it appears that the conventional derivation \cite{AG&D} of the theorem, and thus the result $I_L =0$, would hold as long as the skeleton expansion converges (uniformly, or at least in a point-wise Lebesgue's dominated convergence fashion\footnote{Since each order of the skeleton expansion is a polynomial in $G$, given by absolutely convergent frequency integrals, the chain rule in \eqn{LT invariance T=0} is legitimate order by order, which gives
\be
\int_{-\infty}^\infty \fract{d\ep}{2\pi}\,\Tr\Bigg(\Sigma^{[n]}(i\ep)\;\fract{\partial G(i\ep)}{\partial i\ep}\Bigg)=0\,,\qquad \forall\, n\,,
\label{LT order by order}
\ee
where $\Sigma^{[n]}(i\ep)$ collects all skeleton diagrams of order $n$, evaluated at the exact $G$. Being linear in $\Sigma$, \eqn{LT order by order} holds for every partial sum $S_N(i\ep)\coloneq\sum_{n\leq N}\Sigma^{[n]}(i\ep)$. Whether it survives the limit $N\to\infty$ is therefore a question of exchanging this limit with the frequency integral. By dominated convergence the exchange is justified, and thus $I_L=-\lim_{N\to\infty}\int\frac{d\ep}{2\pi}\Tr\big(S_N\,\partial_{i\ep}G\big)=0$, if $S_N(i\ep)$ converges to the physical $\Sigma(i\ep)$ and there is an integrable function $F(\ep)$ such that $\big|\Tr\big(S_N(i\ep)\,\partial_{i\ep}G(i\ep)\big)\big|\leq F(\ep)$ for all $N$. 
Conversely, wherever $I_L\neq0$ the skeleton series of $\Sigma$ at the exact $G$ either diverges, is not dominated, or converges to an unphysical self-energy~\cite{Kozik2015,Rossi2015,Gunnarsson2017,Essl2026}. Since the examples of Sec.~\ref{A few relevant examples} show $I_L\neq0$ in regions adiabatically connected to the non-interacting ground state, the skeleton expansion can break down already there.}) to the physical solutions.\\

\noindent 
Correspondingly, the Ishikawa-Matsuyama-Volovik topological invariant in two-dimensional insulators can be written, using the same notations as in Sec.~\ref{Sec3}, Eq.~\eqn{IL(1) dream}, as 
\beal
\nu_\text{H\,L} &= -\fract{2\pi c}{e V}\;\fract{\partial N_L}{\partial B}\\
&=  -i\,\fract{\pi}{3}\;\fract{1}{V}\,\sum_\bk\,\int\fract{d\ep}{2\pi}\,\ep_{\mu\nu\rho}\,
\Tr\Big(G(i\ep,\bk)\cdot\partial_\mu G(i\ep,\bk)^{-1}\cdot
\partial_{\nu} G(i\ep,\bk)\cdot\partial_\rho G(i\ep,\bk)^{-1}\Big)\,,
\label{Matsuyama-Ishikawa-Volovik topological invariant}
\eal
and represents the winding number of the map $(\ep,\bk)\to G(i\ep,\bk)$, which is quantized in integer values. However, the physical Hall number is actually 
$\nu_\text{H} = \nu_\text{H\,L} + \nu^I_\text{H\,L}$, where the interaction correction reads, 
through Eq.~\eqn{IL(1) dream}, 
\beal
\nu^I_{H\,L} &= -\fract{2\pi c}{e V}\;\fract{\partial I_L}{\partial B}\\
&= -i\,\fract{\pi}{3}\,\fract{1}{V^2}\,\sum_{\bk\bkp}\, \int\fract{d\ep}{2\pi}\,\fract{d\ep'}{2\pi}\,\ep_{\mu\nu\rho}\,
\Tr\Big(\partial_{\mu} G(i\ep,\bk) \cdot \gamma^A_{0\nu}(i\ep,\bk;i\ep',\bkp)\tdot  \partial_\rho G(i\ep',\bkp)\Big)\,.
\label{topological invariant correction}
\eal
Here, $\gamma^A_{0\nu}$ are 
the derivatives of the component $\Gamma^A_0$ of the irreducible vertex 
antisymmetric under the exchange of the incoming particle-hole pair with the outgoing one, see Eq.~\eqn{Gamma S and A}, taken  
with respect to the transferred frequency ($\partial_0\equiv\partial_{i\omega}$) or the transferred 
momentum ($\partial_1\equiv\partial_{q_x}$ and $\partial_2\equiv\partial_{q_y}$), evaluated 
at $\omega=q=0$. Since $\gamma^A_{0\nu}$ are very peculiar higher order functional derivatives 
of $\Phi[G]$, we could not find a simple interpretation of \eqn{topological invariant correction} comparable to the ones we found for Eqs.~\eqn{Luttinger count}, \eqn{Luttinger integral} and \eqn{Matsuyama-Ishikawa-Volovik topological invariant}. Nonetheless, we can certainly state that
$\nu^I_{H\,L}$ is also invariant under any non-singular variation of the Green's function and that, when Luttinger's theorem is violated and $\nu_\text{H} \not= \nu_\text{H\,L}$, it can assume integer or fractional values.\\
All these insights open up promising avenues for exploring how such anomalies manifest in other topological invariants governing strongly interacting quantum matter.


\section*{Acknowledgements}
We are extremely grateful to Ivan Pasqua, Frank Ebel, Andrey Chubukov, Evgeny Kozik, Georg Rohringer, Anton Markov, Antoine Georges, Nathan Goldman and Lucila Peralta Gavensky for helpful discussions and comments. Benchmarks and checks have been performed with the package \emph{Fermions.jl} \cite{Wallerberger2022}.



\paragraph{Funding information}
HE and AT acknowledge support from Austrian Science Fund (FWF), through projects with 
Grant-DOI 10.55776/I5487 and Grant-DOI 
10.55776/KIN2563725.

\begin{appendix}

\section{Ward-Takahashi identity}
\label{Ward-Takahashi identity}

Since the identity \eqn{Important 4 bis} may not be familiar to everyone, we derive it in this Appendix. \\
The density operator in a basis of Bloch waves $\esp{i\bk\cdot\br}\,u_{n\bk}(\br)$ is
\bealn
\rho(\bq) &= \sum_{nm\bk\sigma}\, \rho_{0\,nm}(\bk,\bk+\bq)\, c^\dagger_{n\bk\sigma}\,
c^\dagga_{m\bk+ \bq\sigma}\,,
\eal
where
\bealn
\rho_{0\,nm}(\bk,\bk+\bq) = \int d\br\, u_{n\bk}(\br)^*\; 
u_{m\bk+\bq}(\br) = \rho_{0,mn}(\bk+\bq,\bk)^*\,.
\eal
For $\bq=0$, $\rho_{0\,nm}(\bk,\bk)=\delta_{nm}$. Up to second order in $\bq$, 
the matrix $\rho_0(\bk,\bk+\bq)=\rho_0(\bk+\bq,\bk)^\dagger$ with elements $\rho_{0\,nm}(\bk,\bk+\bq)$ reads
\be
\rho_0(\bk,\bk+\bq)
= 1 -i\bq\cdot \bd{A}(\bk) + \fract{1}{2}\,\sum_{\mu\nu}\,q_\mu\,q_\nu\,K^{\mu\nu}(\bk)
\,,
\label{App A: density}
\ee
where $\bd{A}(\bk)$ is the Berry connection with elements $\bd{A}_{nm}(\bk)=i\,\bra{u_{n\bk}}\bd{\nabla}_\bk u_{m\bk}\rangle$, while 
\beal
K^{\mu\nu}(\bk) &= -g^{\mu\nu}(\bk) -\fract{i}{2}\,\Big(
\partial_\mu A^\nu(\bk) + \partial_\nu A^\mu(\bk)\Big)
\eal 
with $g^{\mu\nu}(\bk)$ the hermitian quantum metric with elements
\bealn
g^{\mu\nu}_{nm}(\bk) &= \fract{1}{2}\,\Big( 
\bra{\partial_{k_\mu} u_{n\bk}}\partial_{k_\nu} u_{m\bk}\rangle
+ \bra{\partial_{k_\nu} u_{n\bk}}\partial_{k_\mu} u_{m\bk}\rangle\Big)\,.
\eal
For convenience, we rewrite $\rho_0(\bk,\bk+\bq)$ in terms of two hermitian matrices as  
\beal
\rho_0(\bk,\bk+\bq) &= \rho_0(\bk,\bq) - i\,\rho_1(\bk,\bq)\,,& 
\rho_0(\bk+\bq,\bk) &= \rho_0(\bk,\bq) + i\,\rho_1(\bk,\bq)\,,
\label{App A: density separated}
\eal
where $\rho_1(\bk,\bq)$ contains the Berry connection and vanishes at $\bq=\bd{0}$. \\

\noindent
The Ward-Takahashi identity for the charge can be written in matrix notation, see Sec.~4.10 in \cite{libro}, either as 
\beal
&\rho_0(\bk,\bk+\bq)\cdot\Sigma(i\ep_n+i\omega_m,\bk+\bq)
-\Sigma(i\ep_n,\bk)\cdot\rho_0(\bk,\bk+\bq)\\
&\qquad = \fract{1}{V}\,\sum_{\bkp}\,T\sum_{n'}\,
\Gamma_0(i\ep_n,\bk;i\ep_{n'}+i\omega_m,\bkp+\bq|i\ep_{n'},\bkp;i\ep_n+i\omega_m,\bk+\bq)\tdot
\\
&\qquad\qquad\qquad\qquad  \Big(\rho_0(\bkp,\bkp+\bq)\cdot G(i\ep_{n'}+i\omega_m,\bkp+\bq)-
G(i\ep_{n'},\bkp)\cdot\rho_0(\bkp,\bkp+\bq)\Big) \,,
\label{WT-final}
\eal
where $\ep_n=(2n+1)\pi T$ and $\ep_{n'}=(2n'+1)\pi T$ are fermionic Matsubara frequencies, while $\omega_m=2m\pi T$ a bosonic one, or as 
\beal
&\Sigma(i\ep_n+i\omega_m,\bk+\bq)\cdot\rho_0(\bk+\bq,\bk)
-\rho_0(\bk+\bq,\bk)\cdot\Sigma(i\ep_n,\bk)\\
&\qquad = \fract{1}{V}\,\sum_{\bkp}\,T\sum_{n'}\,
\Gamma_0(i\ep_n+i\omega_m,\bk+\bq;i\ep_{n'},\bkp|i\ep_{n'}+i\omega_m,\bkp+\bq;i\ep_n,\bk)\tdot
\\
&\qquad\qquad\qquad\qquad  \Big( G(i\ep_{n'}+i\omega_m,\bkp+\bq)\cdot\rho_0(\bkp+\bq,\bkp)
-\rho_0(\bkp+\bq,\bkp)\cdot G(i\ep_{n'},\bkp) \Big) \,,
\label{WT-finale}
\eal
where $\rho_0(\bk+\bq,\bk)=\rho_0(\bk,\bk+\bq)^\dagger$.\\
If we take half the sum 
of \eqn{WT-final} and \eqn{WT-finale} and use the conventions in Sec.~\ref{Sec3}, we obtain 
\beal
&\fract{1}{2}\,\Big\{\Sigma(i\ep_n+i\omega_m,\bk+\bq),\rho_0(\bk,\bq)\Big\}
-\fract{1}{2}\,\Big\{\Sigma(i\ep_n,\bk),\rho_0(\bk,\bq)\Big\}\\
&\quad +\fract{i}{2}\,\Big[\Sigma(i\ep_n+i\omega_m,\bk+\bq),\rho_1(\bk,\bq)\Big] +\fract{i}{2}\,\Big[\Sigma(i\ep_n,\bk),\rho_1(\bk,\bq)\Big]
\\
&\qquad = \fract{1}{2V}\,\sum_{\bkp}\,T\sum_{n'}\,
\Gamma^S_0(i\ep_n,\bk;i\ep_{n'},\bkp|i\omega_m,\bq)\tdot
\\
&\qquad \qquad  \bigg( \Big\{G(i\ep_{n'}+i\omega_m,\bkp+\bq),\rho_0(\bkp,\bq)\Big\}
-\Big\{G(i\ep_{n'},\bkp),\rho_0(\bkp,\bq)\Big\} \\
&\qquad\qquad \qquad\quad
+i\,\Big[G(i\ep_{n'}+i\omega_m,\bkp+\bq),\rho_1(\bkp,\bq)\Big]
+i\,\Big[G(i\ep_{n'},\bkp),\rho_1(\bkp,\bq)\Big]
\bigg) \\
&\qquad \quad + \fract{1}{2V}\,\sum_{\bkp}\,T\sum_{n'}\,
\Gamma^A_0(i\ep_n,\bk;i\ep_{n'},\bkp|i\omega_m,\bq)\tdot
\\
&\qquad \qquad \qquad  \bigg( \Big[G(i\ep_{n'}+i\omega_m,\bkp+\bq),\rho_0(\bkp,\bq)\Big]
+\Big[G(i\ep_{n'},\bkp),\rho_0(\bkp,\bq)\Big] \\
&\qquad\qquad \qquad\quad
+i\,\Big\{G(i\ep_{n'}+i\omega_m,\bkp+\bq),\rho_1(\bkp,\bq)\Big\}
-i\,\Big\{G(i\ep_{n'},\bkp),\rho_1(\bkp,\bq)\Big\}
\bigg)\,.
\label{symmetric}
\eal
Similarly, the half difference 
of \eqn{WT-finale} and \eqn{WT-final} reads
\beal
&\fract{1}{2}\,\Big[\Sigma(i\ep_n+i\omega_m,\bk+\bq),\rho_0(\bk,\bq)\Big]
+\fract{1}{2}\,\Big[\Sigma(i\ep_n,\bk),\rho_0(\bk,\bq)\Big]\\
&\quad +\fract{i}{2}\,\Big\{\Sigma(i\ep_n+i\omega_m,\bk+\bq),\rho_1(\bk,\bq)\Big\}
-\fract{i}{2}\,\Big\{\Sigma(i\ep_n,\bk),\rho_1(\bk,\bq)\Big\}
\\
&\qquad = \fract{1}{2V}\,\sum_{\bkp}\,T\sum_{n'}\,
\Gamma^A_0(i\ep_n,\bk;i\ep_{n'},\bkp|i\omega_m,\bq)\tdot
\\
&\qquad \qquad \qquad \bigg( \Big\{G(i\ep_{n'}+i\omega_m,\bkp+\bq),\rho_0(\bkp,\bq)\Big\}
-\Big\{G(i\ep_{n'},\bkp),\rho_0(\bkp,\bq)\Big\} \\
&\qquad \qquad \qquad \quad 
+i\,\Big[G(i\ep_{n'}+i\omega_m,\bkp+\bq),\rho_1(\bkp,\bq)\Big]
+i\,\Big[G(i\ep_{n'},\bkp),\rho_1(\bkp,\bq)\Big]
\bigg) \\
&\qquad\quad + \fract{1}{2V}\,\sum_{\bkp}\,T\sum_{n'}\,
\Gamma^S_0(i\ep_n,\bk;i\ep_{n'},\bkp|i\omega_m,\bq)\tdot
\\
&\qquad \qquad  \qquad \qquad \bigg( \Big[G(i\ep_{n'}+i\omega_m,\bkp+\bq),\rho_0(\bkp,\bq)\Big]
+\Big[G(i\ep_{n'},\bkp),\rho_0(\bkp,\bq)\Big] \\
&\qquad \qquad \qquad \quad
+i\,\Big\{G(i\ep_{n'}+i\omega_m,\bkp+\bq),\rho_1(\bkp,\bq)\Big\}
-i\,\Big\{G(i\ep_{n'},\bkp),\rho_1(\bkp,\bq)\Big\}
\bigg)\,.
\label{antisymmetric}
\eal 
One can now take the limit $T\to 0$, so that $\ep_n\to \ep$, $\ep_{n'}\to \ep'$ 
as well as $\omega_m\to \omega$ all become continuous variables, and the sum over 
$\ep_{n'}$ an integral. Subsequently, we   
differentiate both sides of the equations \eqn{symmetric} and \eqn{antisymmetric} with respect to $\bq$ and $\omega$, evaluated at $\omega=q=0$. This process results in a series of identities that relate the derivatives of the self-energy, Green's function, 
and irreducible vertices. \\

\noindent
The results in Sec.~\ref{Sec3} were obtained using a topologically trivial basis where the Berry connection is zero throughout the Brillouin zone, including its boundaries. Therefore, we derive some of those identities in the case where $\bd{A}(\bk)=0$. The simplest ones are obtained by taking the first derivatives of \eqn{symmetric} with respect to $\omega$ or $\bq$, which leads to   
\beal
\partial_{i\ep} \Sigma(i\ep,\bk) &= \fract{1}{V}\,\sum_{\bkp}\,\int\fract{d\ep'}{2\pi}\, 
\Gamma^S_0(i\ep,\bk;i\ep',\bkp)\tdot
\partial_{i\ep'} G(i\ep',\bkp) \,,\\
\partial_{\mu} \Sigma(i\ep,\bk) &= \fract{1}{V}\,\sum_{\bkp}\,\int\fract{d\ep'}{2\pi}\, 
\Gamma^S_0(i\ep,\bk;i\ep',\bkp)\tdot
\partial_{\mu} G(i\ep',\bkp)\,,
\label{simple WT identity}
\eal
where $\partial_{\mu} \Sigma(i\ep,\bk)=\partial_{k_\mu} \Sigma(i\ep,\bk)$ 
and $\partial_{\mu} G(i\ep',\bkp)=\partial_{k'_\mu} G(i\ep',\bkp)$. 
Equation \eqn{Important 4 bis} is instead obtained by the mixed partial derivative of \eqn{antisymmetric} with respect to $\omega$ and $q_\mu$, which yields
\beal
&\fract{1}{V}\,\sum_{\bkp}\,\int\fract{d\ep'}{2\pi}\,
\dot{\Gamma}^A_0(i\ep,\bk;i\ep',\bkp)\tdot\partial_\mu G(i\ep',\bkp)\\
&\qquad =-\fract{1}{V}\,\sum_{\bkp}\,\int\fract{d\ep'}{2\pi}\,
\Gamma^A_{0\mu}(i\ep,\bk;i\ep',\bkp)\tdot\partial_{i\ep'} G(i\ep',\bkp)\,,
\label{non trivial WT identity}
\eal
from which \eqn{Important 4 bis} directly follows.

\section{Invariance property of $\nu^I_{\text{H}\,L}$}
\label{Invariance property}

In this appendix we show that also 
\be
\nu^I_{\text{H}\,L} 
= -i\,\fract{\pi}{3V^2}\sum_{\bk_1\bk_2}\, \int\fract{d\ep_1}{2\pi}\,\fract{d\ep_2}{2\pi}\,\ep_{\mu\nu\rho}\,
\Tr\Big(\partial_{\mu} G(i\ep_1,\bk_1) \cdot \gamma^A_{0\nu}(i\ep_1,\bk_1;i\ep_2,\bk_2)\tdot  \partial_\rho G(i\ep_2,\bk_2)\Big)\,,
\label{App B 1}
\ee
is invariant under any non-singular variation of the Green's function. 
To simplify the notation, we define $\bx_i\coloneq (i\ep_i,\bk_i)$,  
\bealn
\int d\bx_i \;\dots \; \coloneq  \fract{1}{V}\,\sum_{\bk_i}\, \int \fract{d\ep_i}{2\pi}\;\dots\; \,, 
\eal
and further assume that the trace contracts the orbital indices attached to the same $\bx_i$, so that 
\be
\nu^I_{\text{H}\,L} 
= -i\,\fract{\pi}{3} \int d\bx_1\,d\bx_2\; \ep_{\mu\nu\rho}\,
\Tr\Big(\partial_{\mu} G(\bx_1) \, \gamma^A_{0\nu}(\bx_1;\bx_2)\,  \partial_\rho G(\bx_2)\Big)\,.
\label{App B 2}
\ee
We now take a variation and, exploiting the properties of the fully antisymmetric tensor and assuming that integration by parts does not yield any boundary term, we find 
\bealn
\delta\nu^I_{\text{H}\,L} 
&= i\,\fract{2\pi}{3} \int d\bx_1\,d\bx_2\, \ep_{\mu\nu\rho}\,
\Tr\Big( \delta G(\bx_1) \; \partial_{x_{1\mu}}\gamma^A_{0\nu}(\bx_1;\bx_2)\;  \partial_{x_{2\rho}} G(\bx_2)\Big)\\
&\qquad -i\,\fract{\pi}{3} \int d\bx_1\,d\bx_2\, \ep_{\mu\nu\rho}\,
\Tr\Big(\partial_{x_{1\mu}}  G(\bx_1) \; \delta \gamma^A_{0\nu}(\bx_1;\bx_2)\; \partial_{x_{2\rho}}   G(\bx_2)\Big)\\
& = i\,\fract{2\pi}{3} \int d\bx_1\,d\bx_2\, \ep_{\mu\nu\rho}\,
\Tr\Big( \delta G(\bx_1) \; \partial_{\mu}\gamma^A_{0\nu}(\bx_1;\bx_2)\;  \partial_{\rho} G(\bx_2)\Big)\\
&\qquad -i\,\fract{\pi}{3} \int d\bx_1\,d\bx_2\, \ep_{\mu\nu\rho}\,
\Tr\Big(\partial_{\mu}  G(\bx_1) \; \delta \gamma^A_{0\nu}(\bx_1;\bx_2)\; \partial_{\rho}  G(\bx_2)\Big)\,,
\eal
having defined 
\be\nonumber
\partial_{\mu}\gamma^A_{0\nu}(\bx_1;\bx_2) \coloneq \sum_{i=1}^2\,\partial_{x_{i\mu}}\gamma^A_{0\nu}(\bx_1;\bx_2)\,.
\ee
We can write
\beal
\partial_{\mu}\gamma^A_{0\nu}(\bx_1;\bx_2) &= \int d\bx_3\, S_\nu(\bx_1,\bx_2;\bx_3)\,
\partial_\mu G(\bx_3)\,,\\
\delta\gamma^A_{0\nu}(\bx_1;\bx_2) &= \int d\bx_3\, S_\nu(\bx_1,\bx_2;\bx_3)\,
\delta G(\bx_3)\,,
\label{App B def S}
\eal
where $S_\nu(\bx_1,\bx_2;\bx_3)=-S_\nu(\bx_2,\bx_1;\bx_3) \coloneq \delta \gamma^A_{0\nu}(\bx_1;\bx_2)/\delta G(\bx_3)$ is a six-leg vertex. 
Since internal lines do not enjoy the same crossing relation 
$\gamma^A_{0\nu}(\bx_1;\bx_2) = -\gamma^A_{0\nu}(\bx_2;\bx_1)$ of the external legs, the 
six-leg vertices $S_\nu(\bx_1,\bx_2;\bx_3)$, $S_\nu(\bx_3,\bx_2;\bx_1)$ and $S_\nu(\bx_1,\bx_3;\bx_2)$ cannot be mapped onto each other by any simple sign change. However, one can show that 
$S_\nu(\bx_1,\bx_2;\bx_3)$ satisfies the cyclic condition:
\be
S_\nu(\bx_1,\bx_2;\bx_3)+S_\nu(\bx_2,\bx_3;\bx_1)+S_\nu(\bx_3,\bx_1;\bx_2) = 0\,.
\label{App B cyclic}
\ee
Indeed, since the irreducible vertex with transferred frequency and momentum $\by=(i\omega,\bq)$ 
is defined as 
\be
\Gamma_0(\bx_1+\by,\bx_2|\bx_2+\by,\bx_1)=\fract{\delta^2\Phi[G]}{\;\delta G(\bx_1;\bx_1+\by)\,
\delta G(\bx_2+\by;\bx_2)\;}\;,
\ee
then 
\beal
S_\nu(\bx_1,\bx_2;\bx_3) &= \fract{1}{2}\,\partial_{y_\nu}\,\Bigg(
\fract{\delta^3\Phi[G]}{\;\delta G(\bx_1;\bx_1+\by)\,
\delta G(\bx_2+\by;\bx_2)\,\delta G(\bx_3;\bx_3)\;}\\
&\qquad\qquad\qquad
-\fract{\delta^3\Phi[G]}{\;\delta G(\bx_1+\by;\bx_1)\,
\delta G(\bx_2;\bx_2+\by)\,\delta G(\bx_3;\bx_3)\;}\Bigg)_{\big|\by=\bd{0}}\;.
\label{App B explicit S}
\eal
We label each pair $(\bd{a}_i;\bd{b}_i)$ in $\delta G(\bd{a}_i;\bd{b}_i)$, $i=1,2,3$, by its mean
$(\bd{a}_i+\bd{b}_i)/2$ and its transfer $\bd{a}_i-\bd{b}_i=\by\,s_i$, $s_i=0,\pm1$. The $\by$-derivative in
\eqn{App B explicit S} is then the sum of the derivatives with respect to the $\by$-dependent
means and to the $\by$-dependent transfers. The two terms in \eqn{App B explicit S} have the same
means, $(\bx_1+\by/2,\,\bx_2+\by/2,\,\bx_3)$, so the contributions of the shifts of the means
cancel; on the contrary, they have opposite transfers, $(s_1,s_2,s_3)=\pm(-1,1,0)$, so those contributions add.
Hence $S_\nu(\bx_1,\bx_2;\bx_3)$ is the derivative of $\Phi^{(3)}$ at zero transfer along the
direction $(s_1,s_2,s_3)=(-1,1,0)$ in transfer space, the components referring to the pairs attached to $\bx_1$, $\bx_2$ and $\bx_3$, in this order. Since $\Phi^{(3)}$, being a third functional
derivative, is symmetric under permutations of its three pairs, the same ordering can be used
for $S_\nu(\bx_2,\bx_3;\bx_1)$ and $S_\nu(\bx_3,\bx_1;\bx_2)$, which are then the derivatives of
the same function along $(0,-1,1)$ and $(1,0,-1)$. The three directions add up to zero, i.e. there is no shift at linear order in $\by$, hence
the identity \eqn{App B cyclic}.\\

\noindent
Through \eqn{App B def S} we can compactly rewrite $\delta\nu^I_{H\,L}$ as 
\beal
\delta\nu^I_{\text{H}\,L} 
&= i\,\fract{\pi}{3} \int \prod_{i=1}^3\,d\bx_i\, \sum_\mu\,
\Tr\Big( \delta G(\bx_1) \;I_\mu(\bx_2,\bx_3)\; Q_\mu(\bx_1,\bx_2,\bx_3)  
\Big)\,,
\label{App B 3}
\eal
having defined 
\beal
I_\mu(\bx_2,\bx_3) &\coloneq \sum_{\nu\rho}\, \ep_{\mu\nu\rho}\, \partial_{\nu} G(\bx_2)\, 
\partial_\rho G(\bx_3)\,,\\
Q_\mu(\bx_1,\bx_2,\bx_3) &\coloneq 2S_\mu(\bx_1,\bx_2;\bx_3) - S_\mu(\bx_3,\bx_2;\bx_1)
= 2S_\mu(\bx_1,\bx_2;\bx_3) + S_\mu(\bx_2,\bx_3;\bx_1)
\,.
\label{App B 3 bis}
\eal
Since $I_\mu(\bx_2,\bx_3)=-I_\mu(\bx_3,\bx_2)$, we can replace $Q_\mu(\bx_1,\bx_2,\bx_3)$ 
with 
\beal
Q_\mu(\bx_1,\bx_2,\bx_3) &\to \fract{1}{2}\,\Big(
Q_\mu(\bx_1,\bx_2,\bx_3)-Q_\mu(\bx_1,\bx_3,\bx_2)\Big)\\
& = \fract{1}{2}\,\Big(2S_\mu(\bx_1,\bx_2;\bx_3) + S_\mu(\bx_2,\bx_3;\bx_1)
-2S_\mu(\bx_1,\bx_3;\bx_2) - S_\mu(\bx_3,\bx_2;\bx_1)\Big)\\
& = S_\mu(\bx_1,\bx_2;\bx_3) + S_\mu(\bx_2,\bx_3;\bx_1)
+S_\mu(\bx_3,\bx_1;\bx_2)\equiv 0\,, 
\eal
which implies that \eqn{App B 3} vanishes, thus $\delta\nu^I_{H\,L}=0$ for any variation 
$\delta G$ such that the system remains an insulator without a Luttinger surface. The proof ultimately relies on the condition that the Luttinger-Ward functional is three times differentiable, including finite-transfer off-diagonal directions.

\end{appendix}

\bibliography{mybiblio,Bibliography}




\nolinenumbers

\end{document}